\documentclass[9pt,conference]{IEEEtran}
\IEEEoverridecommandlockouts
\newcommand{\secref}[1]{Section~\ref{#1}}
\newcommand{\figref}[1]{Fig.~\ref{#1}}
\newcommand{\tabref}[1]{Table~\ref{#1}}

\usepackage{cite}
\usepackage{siunitx}
\usepackage{tikz}
\usepackage{amsmath,amsthm,amssymb,amsfonts}
\usepackage{algorithm,algorithmic}
\usepackage{graphicx}
\usepackage{textcomp}
\usepackage{xcolor}
\usepackage[caption=false,font=footnotesize,labelfont=rm,textfont=rm]{subfig}
\usepackage{makecell,multirow}
\usepackage{booktabs}
\IEEEaftertitletext{\vspace{-1.8\baselineskip}}

\def\BibTeX{{\rm B\kern-.05em{\sc i\kern-.025em b}\kern-.08em
    T\kern-.1667em\lower.7ex\hbox{E}\kern-.125emX}}

\begin{document}
\title{
ATSim3.5D: A Multiscale Thermal Simulator for 3.5D-IC Systems based on Nonlinear Multigrid Method
}

\author{\IEEEauthorblockN{Qipan Wang}
\IEEEauthorblockA{
    \textit{Academy for Advanced Interdisciplinary Studies} \\
    \textit{School of Integrated Circuits, Peking University} \\ 
    {Beijing, China}\\
    qpwang@pku.edu.cn
}
\and
\IEEEauthorblockN{Tianxiang Zhu}
\IEEEauthorblockA{
    \textit{School of Integrated Circuits} \\ \textit{Peking University} \\
    {Beijing, China}\\
    txzhu@pku.edu.cn
}
\and
\IEEEauthorblockN{Yibo Lin$^\dagger$}
\IEEEauthorblockA{
    \textit{School of Integrated Circuits, Peking University} \\ 
    \textit{Institute of Electronic Design Automation, Wuxi} \\
    {Beijing, China}\\
    yibolin@pku.edu.cn \\ {$^\dagger$\ \footnotesize{Corresponding Author}}
}
\and
\IEEEauthorblockN{Runsheng Wang}
\IEEEauthorblockA{
    \textit{School of Integrated Circuits, Peking University} \\ 
    \textit{Institute of Electronic Design Automation, Wuxi} \\
    {Beijing, China}\\
    r.wang@pku.edu.cn
}
\and
\IEEEauthorblockN{Ru Huang}
\IEEEauthorblockA{
    \textit{School of Integrated Circuits, Peking University} \\ 
    \textit{Institute of Electronic Design Automation, Wuxi} \\
    {Beijing, China}\\
    ruhuang@pku.edu.cn
}
}
\maketitle
\renewcommand{\thefootnote}{\fnsymbol{footnote}}
\footnotetext[1]{{This work was supported in part by the National Science Foundations of China (Grant No. 62125401, 62034007), the Natural Science Foundation of Beijing, China (Grant No. Z230002), Grant QYJS-2023-2303-B, Beijing Outstanding Young Scientist Program (JWZQ20240101004), and the 111 project (B18001).}
}

\begin{abstract}
To resolve the rising temperatures in 3.5D-ICs, a thermal-aware design flow becomes increasingly crucial, necessitating an accurate and efficient thermal simulation tool. However, previous tools struggle to handle the unique heterogeneous multiscale structures in 3.5D-ICs and the nonlinear thermal effects caused by high temperatures.
In this work, we present a multiscale thermal simulator for 3.5D-ICs. We propose a hybrid tree structure to generate multilevel grids and capture the multiscale features and employ the nonlinear multigrid method for quick solving. 
Compared to ANSYS Icepak, it exhibits high accuracy (mean absolute relative error $<1\%$, max error $<\SI{2}{\degreeCelsius}$), and efficiency ($80\times$ acceleration), delivering a powerful means to evaluate and refine thermal designs.
\end{abstract}


\section{Introduction}
Over the past years, integrated circuits have become gradually hotter due to aggressive integration in the pursuit of high performance. 
The situation gets aggravated in advanced packaging, including 2.5D/3D/3.5D-ICs, which improve performance in small footprints by integrating powerful dies close to each other. 
To resolve thermal issues, various approaches have been proposed, ranging from cooling techniques to thermal-aware design and management \cite{shukla2019overview,2024atplace}, all relying on accurate thermal simulation.

Many thermal simulation tools have been available \cite{ansys,comsol,stan2003hotspot,terraneo20213d,ladenheim2018mta,wang2024atsim3d,2024fastherm}, most of which are based on numerical methods, such as COMSOL \cite{comsol}, MTA \cite{ladenheim2018mta} (finite element method), ANSYS Icepak \cite{ansys}, ATSim3D \cite{wang2024atsim3d} (finite volume method, FVM), HotSpot \cite{stan2003hotspot}, 3D-ICE \cite{terraneo20213d} (finite difference method). 
It is well known that these discretization-based methods involve a trade-off between accuracy and efficiency, depending on the system size and mesh resolution.
As modern chips involve multiscale features and are evolving towards larger sizes, directly discretizing and solving the problem becomes increasingly difficult or even impossible. 
Therefore, new challenges arise concerning fast and accurate thermal simulation for 3.5D-ICs. 

This work focuses on the 3.5D-ICs\cite{10195617,lau2025current}, a fusion of 2.5D and 3D-ICs, composed of multiple chiplets, which can potentially incorporate 3D-IC components. 
The system encompasses features of various scales, from large-scale heat spreaders and sinks to tiny modules, blocks, or through silicon via (TSVs), presenting the first challenge of multiscale simulation.
A multiscale thermal model \cite{10287686} is developed for chiplet systems, but is restricted to structures in the interposer, including TSVs and redistribution layers (RDL), with little focus on internal chiplet modules.
In \cite{yang2006isac}, the multilevel grid was constructed in a spatial and temporal adaptive manner to accelerate the simulation. 
Moreover, a non-uniform quadtree mesh \cite{smy2001transient} is proposed to refine the grids around hotspots. However, these methods assume a structure in which layers are stacked with a consistent thickness, which is not directly applicable to 3.5D-ICs, where chiplets may have different thicknesses and misaligned layers.

\begin{table}[t]
\footnotesize
\centering
\caption{Comparison of common thermal simulators' features. 'HeatSink\&Spreader` represents the ability to handle the pyramid-shaped heat sink and spreader structure.}
\resizebox{0.5\textwidth}{!}{
\begin{tabular}{c|cc|ccc}
\toprule
    Simulator & Algorithm & Efficiency & \makecell[c]{Nonlinear\\Simulation} & \makecell[c]{HeatSink\\\&Spreader} & 3.5D-IC \\ \hline
    COMSOL \cite{comsol} & \multirow{2}{*}{FEM} & Low & \checkmark & \checkmark & \checkmark \\
    MTA \cite{ladenheim2018mta} & & Medium & \checkmark & \checkmark & \checkmark \\ \hline
    HotSpot \cite{stan2003hotspot} & \multirow{2}{*}{FDM} & Low & $\times$ & \checkmark & $\times$ \\
    3D-ICE \cite{terraneo20213d} & & Low & Sink$^\dagger$ & \checkmark & $\times$ \\ \hline
    Icepak \cite{ansys} & \multirow{3}{*}{FVM} & Low & \makecell[c]{Power} & \checkmark & \checkmark \\
    ATSim3D \cite{wang2024atsim3d} &  & High & \checkmark & $\times$ & $\times$ \\
    This work & & {High} & {\checkmark} & \checkmark & \checkmark \\ \bottomrule
\multicolumn{6}{l}{
$^\dagger$\small{3D-ICE focuses on nonlinear sink model only, rather than conductivity and power.}
}
\end{tabular}
}
\label{table:comp}
\end{table}
    
\begin{figure}[tb]
\centering
\includegraphics[width=.95\linewidth]{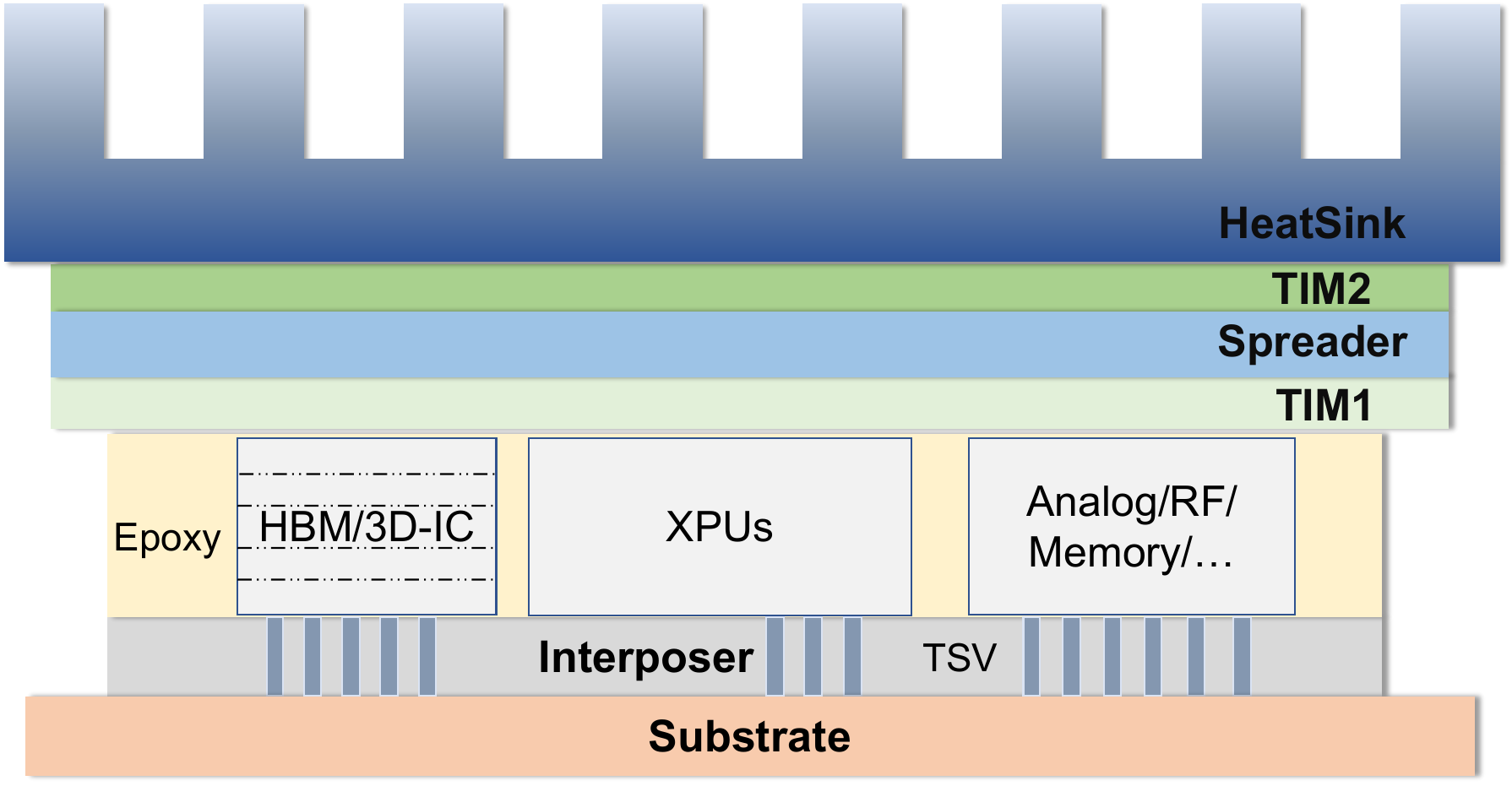}
\caption{Cross section of the 3.5D-IC system with heat sink.}
\label{Fig:25DIC}
\end{figure}

To capture the multiscale features in 3.5D-ICs, the multigrid (MG) method  manifests itself as a powerful tool. Its basic idea is to repeatedly smooth out the residuals from the fine grid to the coarse grid (restriction), subsequently interpolating the correction updates from the coarse grid to the fine grid (prolongation) after hierarchically meshing the system.
However, the high temperature inside 3.5D-ICs hinders the direct application of MG, which may bring about pronounced nonlinear thermal effects and make the governing equation nonlinear as demonstrated by ATSim3D \cite{wang2024atsim3d}, 
To treat the nonlinearity, the full approximation scheme multigrid (FAS-MG)\cite{henson2003multigrid} is a well-established technique, which has yet to be fully explored in the context of thermal simulations within 3.5D-ICs.

In this paper, we present a steady-state thermal simulator \textbf{ATSim3.5D}\footnote{Compiled code \& cases available at: https://github.com/Brilight/ATSim3D\_pub.git} for general multiscale 3.5D-IC systems. We summarize and compare the features of common thermal simulators in \tabref{table:comp}. Our simulator is validated against ANSYS Icepak in two 2.5D/3.5D-IC systems. It exhibits high accuracy (mean absolute relative error $<1\%$, max error $<\SI{2}{\degreeCelsius}$), and efficiency ($80\times$ acceleration in average). 
Our contributions are summarized as follows.
\begin{itemize}
    \item We propose a new multilevel grid generation method to partition the 3.5D-ICs into hybrid quadtrees to capture multiscale features.
    \item We propose a new formulation to calculate the equivalent thermal model for grids with heterogeneous materials.
    \item We propose a FAS-MG-based nonlinear simulator, combined with a customized prolongation and restriction scheme for fast convergence.
\end{itemize}

The rest of this paper is organized as follows. \secref{sec:prelim} reviews the basic of 3.5D-IC thermal model and nonlinear simulation. \secref{sec:algo} thoroughly explain the proposed algorithm. \secref{sec:result} demonstrates the power of our simulator with comprehensive results, followed by the conclusion in \secref{sec:conclu}.

\section{Preliminaries} \label{sec:prelim}
In this section, we first introduce the 3.5D-IC thermal model in \ref{sec:models}. Then we review the basics of the nonlinear multigrid method in \ref{sec:nonlinear}. Finally, we formulate the problem in \ref{sec:metric}.

\subsection{Thermal Models} \label{sec:models}
\subsubsection{3.5D-IC Configuration}
\figref{Fig:25DIC} shows the simplified 3.5D-IC structure.
At the bottom is the substrate, and above it is the interposer layer. TSVs penetrate through the interposer for exterior connections.
A chiplet layer is located above, encompassing various chiplets like XPUs (CPU, GPU, etc.), Memory (DRAM, HBM, etc.), analog or RF devices, 3D-ICs, etc. 
Epoxy molder compound (EMC) fills the space between components inside the chiplet layer, which is coated with thermal interface material (TIM1) to stick to the heat spreader layer. 
A cubic heat sink, which can mimic many realistic situations by changing its convective heat transfer coefficient (HTC), is positioned above the previous layers, bonded with a TIM2 layer.

\subsubsection{Governing Equation} 
In this work, we focus on the steady-state temperature profiles under certain power workloads, a typical concern for thermal-aware design, with the governing nonlinear heat equation reads:
\begin{equation}
    \nabla\cdot\left(\kappa(\textbf{r},T)\nabla T(\textbf{r})\right) = -\textbf{P}(\textbf{r},T),
    \label{Equ:Fourier}
\end{equation}
subject to the convection boundary condition (B.C.) at both top and bottom surfaces (adiabatic for all other surfaces): $-(\kappa\vec{n}\cdot\nabla T)|_{\text{Bottom\&Top}} = h(\textbf{r}, T)(T-T_{amb})$,
where $T(\textbf{r})$ is the 3D temperature $[K]$ profile over the location $\vec{r}=(x,y,z)$. 
$\kappa(\textbf{r},T)$ is the heterogeneous conductivity $[W/m\cdot K]$ that observes the power-law dependence with the temperature \cite{wang2024atsim3d} for silicon: $\kappa(T)= \kappa_0 (T_0/T)^\alpha, \label{equ:kappa}$
where $\kappa_0$ is the conductivity at temperature $T_0$, and $\alpha$ is a power law constant. 
$\textbf{P}(\textbf{r},T)=\textbf{P}_{\text{dyn}}(\textbf{r})+\textbf{P}_{\text{leak}}(\textbf{r},T)$ is the sum of dynamic and leakage power densities $[W/m^3]$. The dynamic power exhibits little dependence on the temperature, while leakage power follows the relation \cite{wang2024atsim3d}:
$\textbf{P}_{\text{leak}}= \textbf{P}_{\text{L0}}\cdot e^{\beta (T-T_0)}, \label{equ:leakage}$
where $\textbf{P}_{\text{L0}}$ is the base leakage power, and $\beta,\ T_0$ the temperature coefficient and base temperature. 

\subsection{Nonlinear Multigrid} \label{sec:nonlinear}
Newton methods are often employed to tackle the nonlinear leakage and conductivity.
\cite{ramalingam2006accurate} proposed to handle the nonlinear conductivity by Newton-Raphson iteration. 
\cite{yan2017efficient} proposed a corrected linearized model for leakage power approximation within algebraic multigrid. 
Their approach belongs to Newton-MG, which, along with FAS-MG, are two common generalizations of MG for handling nonlinear problems. 
The main difference is that the value for FAS-MG to solve on the coarse grids is the exact solution instead of the error (residual) for the linear version. Our simulator leverages the FAS-MG to handle the nonlinear thermal simulation discussed.



\subsection{Problem Formulation} \label{sec:metric} 
This work focuses on solving the nonlinear equation \eqref{Equ:Fourier} under the B.C.s. To measure the accuracy of the simulation results, we define three error metrics: mean absolute error ($\text{MAE} = \textbf{Mean}\{\|T-T_{ref}\|\}$), maximum error ($\text{MaxE} = \textbf{Max}\{\|T-T_{ref}\|\}$), and mean absolute relative error ($\text{MARE} = \textbf{Mean}\{\frac{T-T_{ref}}{T_{ref}-T_\text{amb}}\}$), 
where $T$, $T_{ref}$ are the temperature profile calculated by our simulator and rigorous simulator, and the ambient temperature $T_\text{amb}$ is introduced in \text{MARE} for regularization.
\begin{figure}[tbh]
\centering
    \subfloat[V100 structure]{
    \begin{minipage}[t]{0.28\linewidth}
        \centering
        \includegraphics[width=1\linewidth]{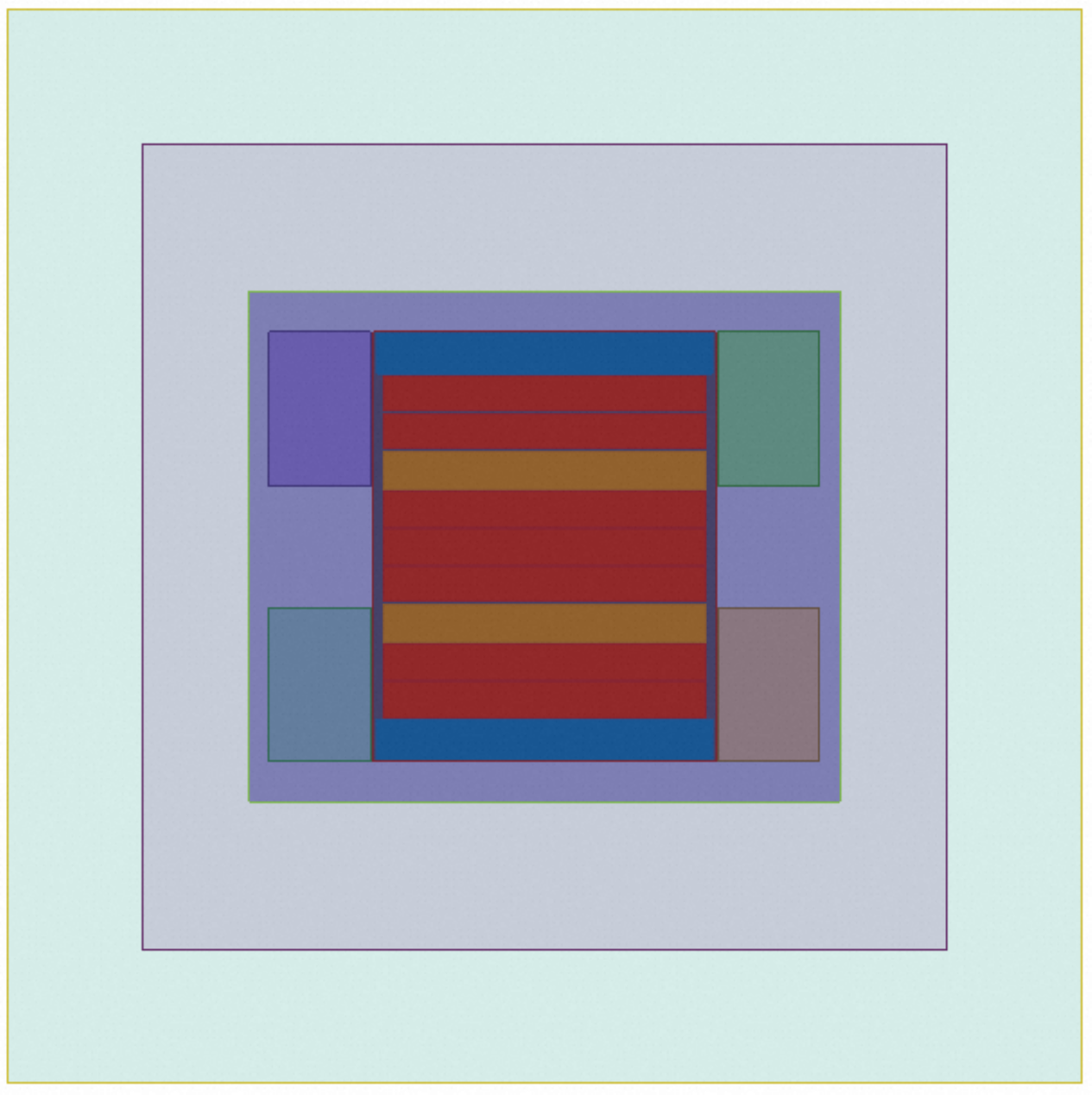}
    \end{minipage}
    }
    \subfloat[Mesh by Icepak]{
    \begin{minipage}[t]{0.29\linewidth}
        \centering
        \includegraphics[width=1\linewidth]{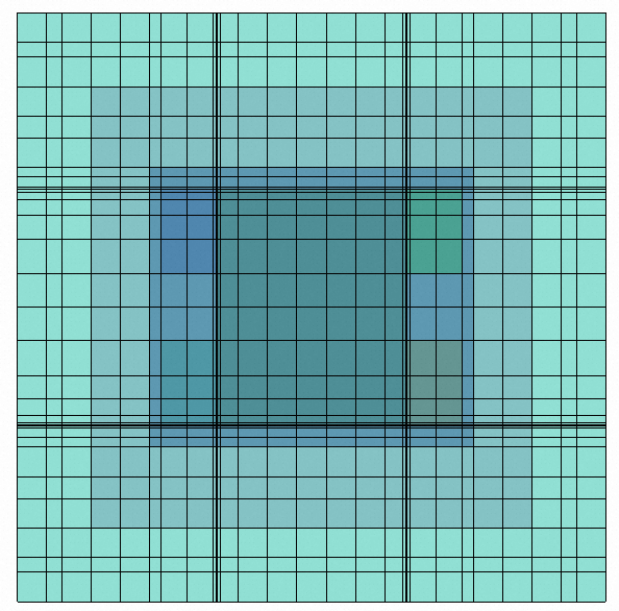}
    \end{minipage}
    }
    \subfloat[Mesh by HotSpot]{
    \begin{minipage}[t]{0.28\linewidth}
        \centering
        \includegraphics[width=1\linewidth]{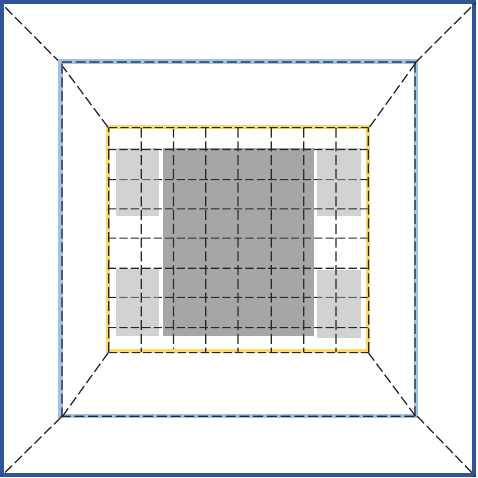}
    \end{minipage}
    }
\caption{Mesh generated for the heat sink layer of the V100 system (a) by different simulators: Icepak (b), HotSpot (c).}
\label{fig:mesh}
\end{figure}

\begin{figure}[tb]
    \centering
    \includegraphics[width=0.98\linewidth]{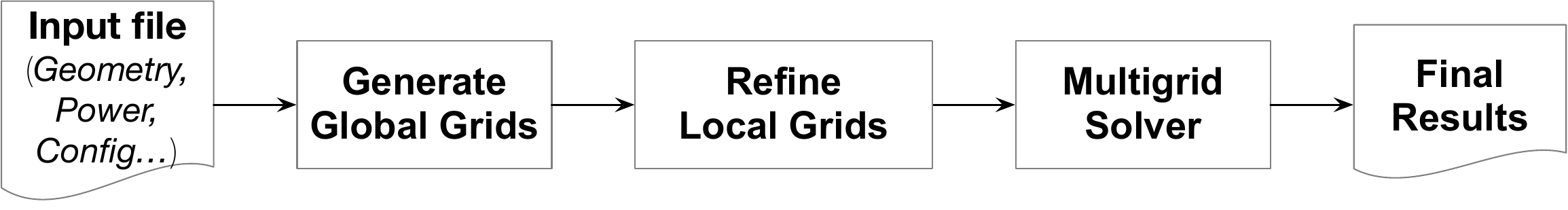}
    \caption{Solving flow of our simulator.}
    \label{Fig:flow}
\end{figure}

\begin{figure*}[tbh]
\centering
    \subfloat[Grids at Heat Sink]{
    \begin{minipage}[t]{0.21\linewidth}
        \centering
        \includegraphics[width=1\linewidth]{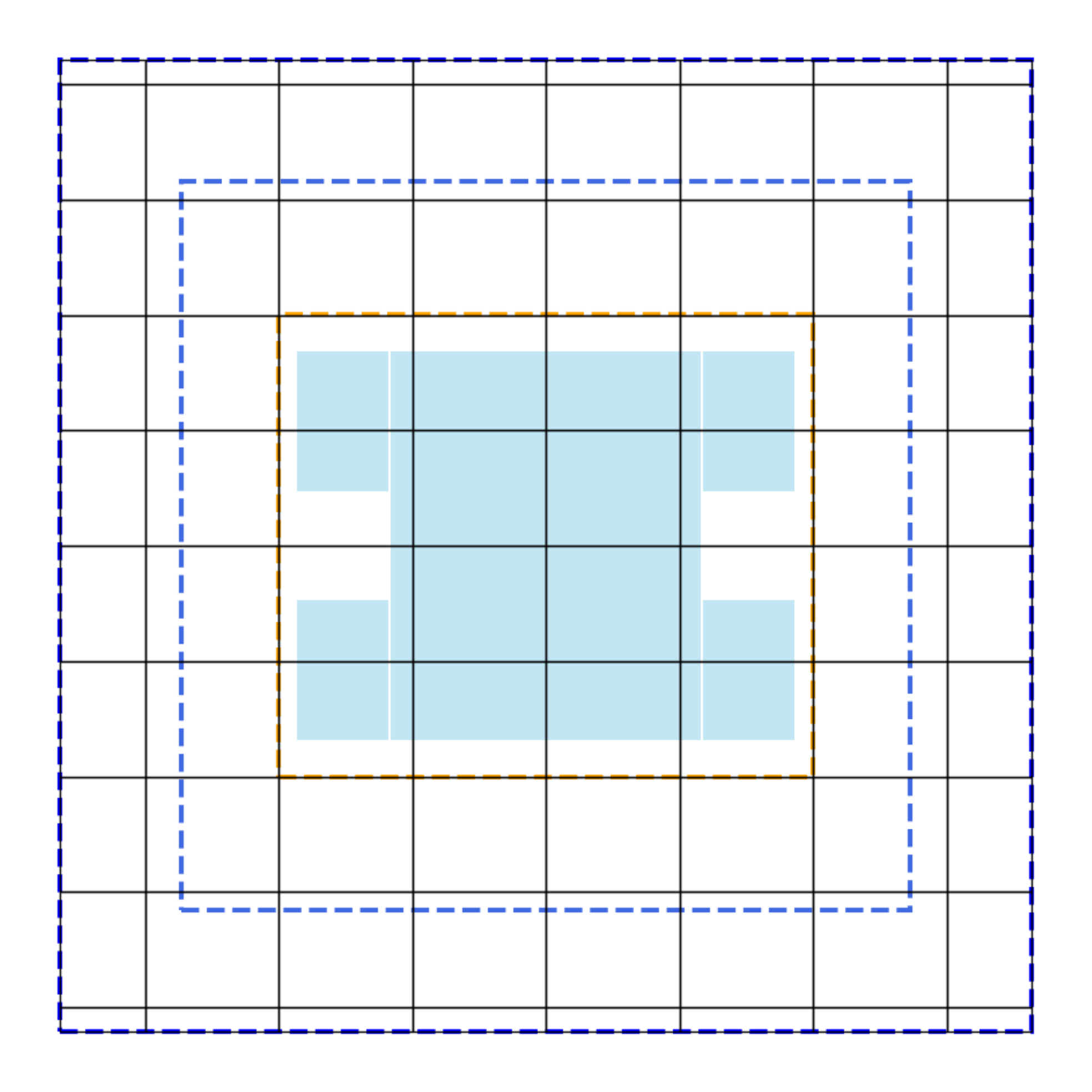}
    \end{minipage}
    }
    \subfloat[Level 1 Grids]{
    \begin{minipage}[t]{0.24\linewidth}
        \centering
        \includegraphics[width=1\linewidth]{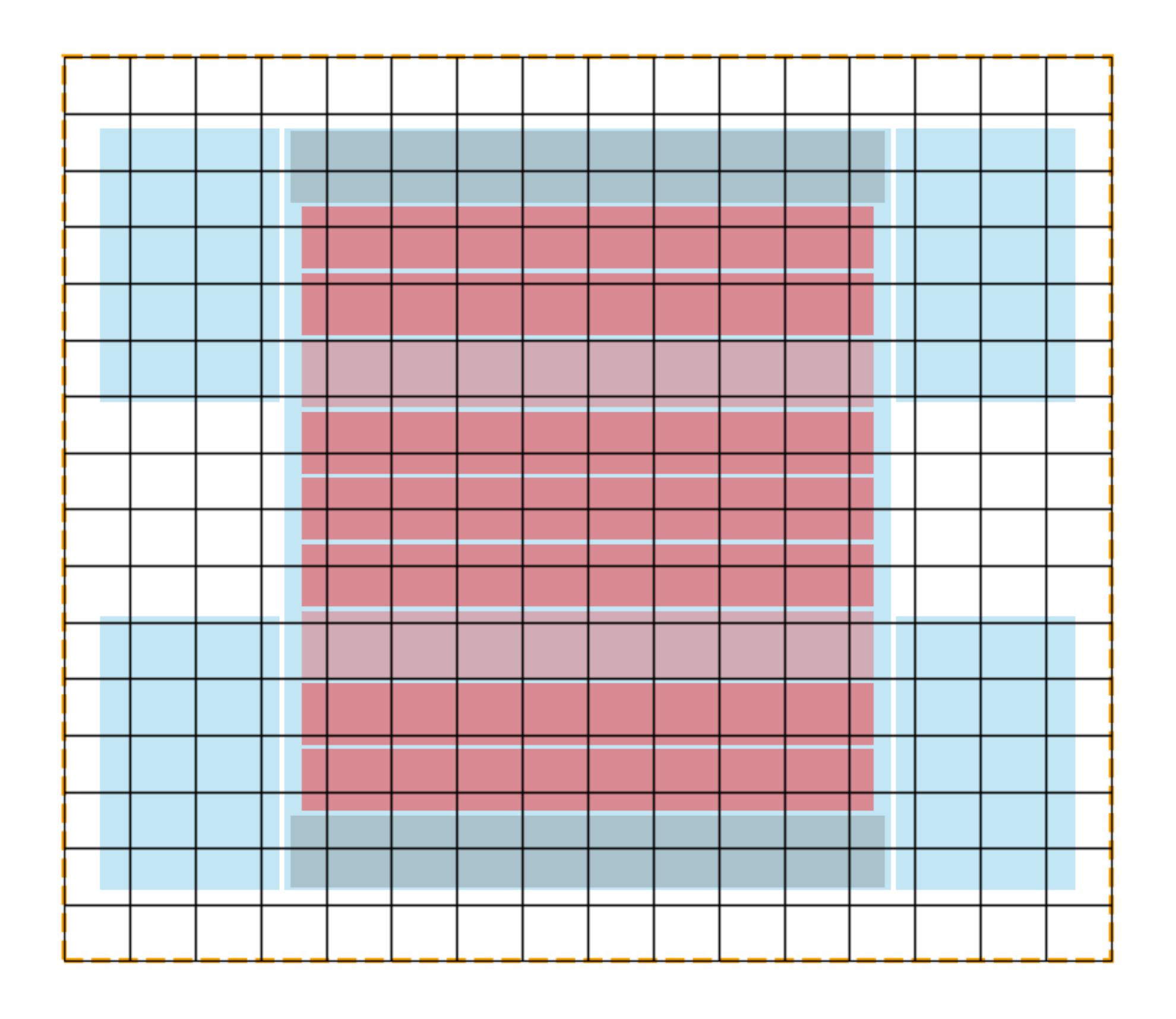}
    \end{minipage}
    }
    \subfloat[Level 2 Grids]{
    \begin{minipage}[t]{0.24\linewidth}
        \centering
        \includegraphics[width=1\linewidth]{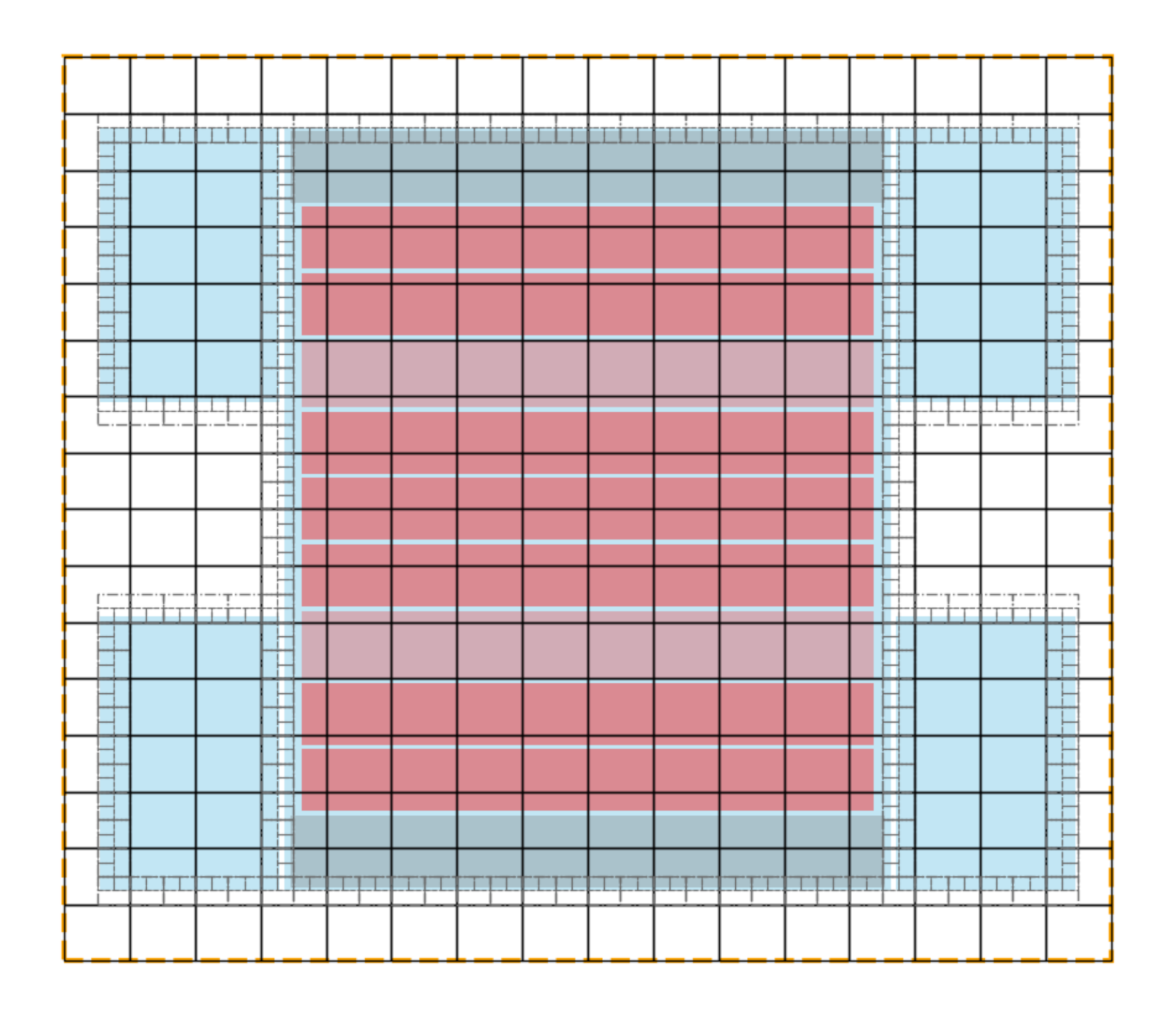}
    \end{minipage}
    }
    \subfloat[Level 3 Grids]{
    \begin{minipage}[t]{0.24\linewidth}
        \centering
        \includegraphics[width=1\linewidth]{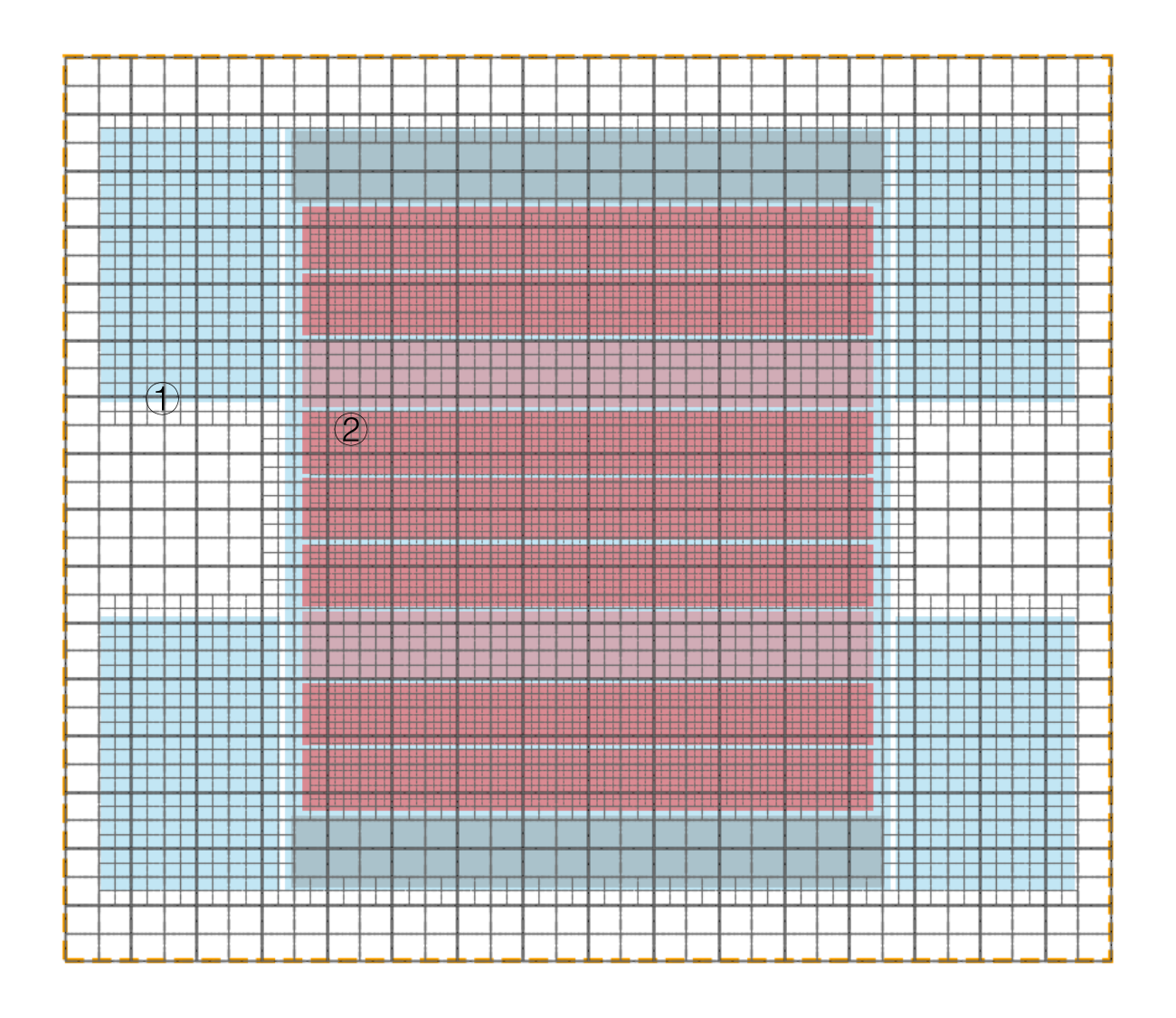}
    \end{minipage}
    }
\caption{Multilevel grids generated for the  heat sink (a) and chiplet layer (b-d).}
\label{fig:atmesh}
\end{figure*}

\section{Whole Framework} \label{sec:algo}
In this section, we illuminate the algorithm of our simulator. We first introduce the motivation in \ref{sec:motiv}, then detail the construction of multilevel grids in \ref{sec:mesh}. The nonlinear simulation algorithm, including FVM formulation and FAS-MG, is presented in \ref{sec:FAS}.

\subsection{Motivation} \label{sec:motiv}
For illustration, consider a simplified Nvidia GV100 GPU \cite{NVIDIATeslaV100}, consisting of a GPU core and four surrounding high-bandwidth memory (HBM) modules as shown in \figref{fig:mesh}(a). 
The GPU core encompasses 84 streaming multiprocessors (SMs), which are consolidated into seven SM groups (SMGs) for simplicity, as denoted by the red rectangles in the figure. 
The core also includes two segments of the L2 cache, illustrated in yellow, along with multiple interface modules, depicted in blue.
\figref{fig:mesh}(b) depicts the nonuniform continuous meshes generated by Icepak for the heat sink layer, and we can observe that fine meshing near localized hotspots extends throughout the system, resulting in excessive meshing. 
In contrast, uniform grids are generated by the influential academic tool HotSpot, as shown in \figref{fig:mesh}(c), for each layer except the blocks at the periphery of the heat spreader and sink. Such an approach leads to over-dense meshing in central regions, 
while the peripheral areas become too sparsely meshed, thus reducing the accuracy. 
To mitigate this issue, \cite{smy2001transient} proposed a rectangular discontinuous mesh that allows local refinement at the locations with steep temperature gradients. However, this does not apply well to 3.5D-IC structures, especially when considering the EMC between chiplets and the unique internal layer structure of each chiplet.

Inspired by the above insights, we propose a multilevel grid generation approach specifically designed for 3.5D-IC systems. With such a multilevel mesh in place, employing a multigrid method to solve the governing nonlinear equation is natural. 
The simulation flow is shown in \figref{Fig:flow}.
Given the input file, including the geometrical configuration, floorplan, and material parameters, the simulator will first preprocess this information. 
Then, it constructs global grids for the whole package system, followed by localized refinement. 
Details on the meshing and solving algorithms will be provided in the coming sections.


\subsection{Multilevel Grid} \label{sec:mesh}
The proposed meshing algorithm starts by partitioning global grids for the whole package, followed by local refinement for both chiplets and other areas of interest, where different chiplets can be independently meshed according to their structural characteristics. Finally, the system will be organized by a hybrid quadtree structure.

\subsubsection{Global Partition}
We propose to construct the global grids layer by layer and treat the layers corresponding to heating and cooling distinctively, as it is essential to tackle the dimension mismatch between heating and cooling components.
The process begins with dividing the chiplet layer into uniform coarse grids, as shown in \figref{fig:atmesh}(b). 
For other layers close to the chiplet layer, such as the interposer, since they often share the same size as the chiplet layer, the global grids are generated in the same manner. 
For cooling components, including heat spreader and heat sink, where temperature variations are relatively mild in both horizontal and vertical directions, it is unnecessary to keep the same grid size.
Therefore, these layers are divided by grids with twice the size of coarse grids. \figref{fig:atmesh}(a) illustrates the grids generated for the heat sink layer. Residual parts in the edge regions smaller than the doubled coarse grid size will remain. In the vertical direction, the uniform partition is conducted for each layer.
Such a partition strategy can avoid not only the imbalance of grid density at the center and edges of the sink in HotSpot but also the continuous fine mesh that causes excessive computational burden in Icepak. 
The temperature-dependent cooling effects can also be simulated in our framework. 

\subsubsection{Local Refinement}
Traditional methods tend to construct equivalent chip thermal models (CTM) in isolated situations for chiplets, replace them in real systems with CTM to reduce the computational burden, and overlook their internal structures.
Nevertheless, since the temperature of the hottest module within the chip is of primary concern, it is crucial to simulate the entire chip-package system and observe the thermal behaviors of all internal modules.
For this purpose, local refinement is required based on the coarse grids generated in the previous stage, utilizing layout information to refine grids into finer ones iteratively.
For instance, chiplets with relatively uniform power distribution can be partitioned coarsely, while structures with irregular power maps, like CPUs and ASICs, should be finely meshed at areas with high power density. For 3D-ICs like HBM, a layer-wise partitioning can be applied. 
Considering different hierarchical structures across various chiplets, we introduce a novel hybrid quadtree approach to control this refinement process.

The refinement process involves two steps. 
In the first step, beginning with coarse grids in the chiplet layer, if certain grids overlap with multiple chiplets, they are split into subgrids along horizontal axes, culminating in a quadtree covering the entire chiplet layer. 
The partition process continues until all leaf nodes of the tree (grids without subgrids) encompass just one chiplet, as depicted in \figref{fig:atmesh} (c). 
Since chiplets are always spaced around hundreds of micrometers apart, the smallest subgrid size will also correspond to this scale.
In the second step, each grid will be further divided in both horizontal and vertical directions, as shown in \figref{fig:atmesh} (d). In the vertical direction, the grid is partitioned according to the number of layers in the chiplet. In horizontal ones, a layer-wise quadtree is generated in the same manner as the first step, except for different partition criteria:
\begin{itemize}
    \item If a grid contains components with not only one material, it will be partitioned. 
    \item If the total power density inside a grid is larger than the given density threshold, which is typically the average power density in our framework, it will be partitioned, since it may have a large thermal gradient inside.
\end{itemize}
These two criteria are demonstrated by the grids at \textcircled{1} and \textcircled{2} in \figref{fig:atmesh} (d), respectively. The partition for all grids stops when their sizes have reached the minimum grid size (solving resolution) or their depths equal the maximum level $L$.

\subsubsection{Hybrid Tree Structure}
After the global partition and local refinement, the whole system can be described within a hybrid tree structure. Different from the hybrid octree by \cite{yang2006isac} and the quadtree by \cite{smy2001transient}, global grids generated in the global partition stage constitute the first layer of the tree, and the number of subgrids for each grid is always a multiple of four, a result of the layer-wise quadtree partition.
To traverse the tree, we adopt the depth-first search, recording the level and information of each grid. For grids with subgrids, their levels are just their depth; while for the others, they will correspond to multiple levels, ranging from their depth to the maximum level $L$.

\subsection{Nonlinear Simulation} \label{sec:FAS}

After generating multilevel grids for the system, we will solve the nonlinear equation Eq.\eqref{Equ:Fourier}. In this section, we will first introduce the equivalent thermal model of compounds, followed by the discretized governing equation according to the FVM principle. Finally, we will present the FAS-MG solver.

\subsubsection{Equivalent Thermal Model}
It is essential to establish a equivalent thermal model, since grids may include multiple materials in 3.5D-ICs, such as epoxy in the chiplet layer or TSVs within the interposer.
Different from previous formulations of arithmetic average or equivalent resistance \cite{wang2023efficient}, here we adopt the Voigt-Reuss bound proposed by \cite{liu2019efficient}, to facilitate the construction of the subsequent restriction operator. The  effective thermal conductivity for a grid reads: 
$\tilde{\kappa} = ((\kappa^R+\kappa^V)/2+\sqrt{\kappa^R\cdot\kappa^V})/2$, 
here $\kappa^V,\kappa^R$ are the arithmetic mean (Voigt approximation) and harmonic mean (Reuss approximation) of all compounds considering their volume ratios in the grid.

\begin{figure}[htbp]
\centering
    \begin{tikzpicture}[scale=.5]
        \large
        \draw[thick] (-6, -3) rectangle (0, 3);
        \draw[thick] (0, -3) rectangle (3, 0);
        \draw[thick] (3, -3) rectangle (6, 0);
        \draw[thick] (0, 0) rectangle (3, 3);
        \draw[thick] (3, 0) rectangle (6, 3);
        
        \node[left] at (-5, 2.2) {$\textbf{p}$};
        \node[circle, draw, fill=black, inner sep=0pt, minimum size=5pt] (T_p) at (-3, 0) {};
        \node[above right] at (-3, 0) {$T_{p}$};

        \node[left] at (1, 2.2) {$\textbf{q}$};
        \node[circle, draw, fill=black, inner sep=0pt, minimum size=5pt] (T_q) at (3, 0) {};
        \node[above right] at (3, 0) {$T_{q}$};
        
        \node[right] at (2, 2.5) {$\textbf{q}_1$};
        \node[circle, draw, fill=black, inner sep=0pt, minimum size=4pt] (T_q1) at (1.5, 1.5) {};
        \node[below right] at (1.5, 1.5) {$T_{q_1}$};

        \node[right] at (2, -0.5) {$\textbf{q}_2$};
        \node[circle, draw, fill=black, inner sep=0pt, minimum size=4pt] (T_q2) at (1.5, -1.5) {};
        \node[below right] at (1.5, -1.5) {$T_{q_2}$};
        
        \node[right] at (5, 2.5) {$\textbf{q}_3$};
        \node[circle, draw, fill=black, inner sep=0pt, minimum size=4pt] (T_q3) at (4.5, 1.5) {};
        \node[below right] at (4.5, 1.5) {$T_{q_3}$};

        \node[right] at (5, -0.5) {$\textbf{q}_4$};
        \node[circle, draw, fill=black, inner sep=0pt, minimum size=4pt] (T_q4) at (4.5, -1.5) {};
        \node[below right] at (4.5, -1.5) {$T_{q_4}$};

        \node[circle, draw, fill=black, inner sep=0pt, minimum size=5pt] (u_sigma_1) at (0, 1.5) {};
        \node[below left] at (0, 1.5) {$T_{\sigma 1}$};
        \node[circle, draw, fill=black, inner sep=0pt, minimum size=5pt] (u_sigma_2) at (0, -1.5) {};
        \node[below left] at (0, -1.5) {$T_{\sigma 2}$};
        
    \end{tikzpicture}
\caption{Notation for adjacent finite volumes $\textbf{p, q}$.}
\label{Fig:Notation}
\end{figure}
    
\subsubsection{FVM Formulation}
We first introduce the necessary notations in \figref{Fig:Notation}. Let $\mathcal{T}$ be the set of all adaptive grids, and $\mathcal{T}_l$ represents the set of grids in level $l$. Every finite control volume $\textbf{p}$ has a solution value $T_p$ constant over $\textbf{p}$, attached to these grids' center.
For any $\textbf{p}\in\mathcal{T}_l$ of measure $\|\textbf{p}\|$ (with size in the x,y,z direction to be $Lx_p,Ly_p,Lz_p$), $\mathcal{N}_l(\textbf{p})$ is the set of all neighbors $\textbf{q}$ in the same level. 
The grid $\textbf{q}$ is divided into four fine grids $\textbf{q}_1\sim\textbf{q}_4\in\mathcal{T}_{l+1}$ further. Here, for simplicity, we assume it is divided only in x and y directions, while the formulation can be generalized to the vertical direction.
Common interface of $\textbf{p}$ and $\textbf{q}$ is a face $\sigma$ with a nonzero measure $\|\sigma\|$ in 2D. $T_{\sigma 1}$ and $T_{\sigma 2}$ represent the unknown values of the interfaces between $\textbf{p}$ and $\textbf{q}_1$, $\textbf{q}_2$, respectively. 

To formulate the discretized governing equation for the grid $\textbf{p}$, integrate the equation \eqref{Equ:Fourier} over its control volume and apply the divergence theorem:     
$\int_{\partial \textbf{p}}\kappa(\textbf{r},T)\nabla T(\textbf{r})\cdot \vec{n}_\textbf{p} ds = -\int_{\textbf{p}}\textbf{P}(\textbf{r},T)=-\mathbf{P}(T_\textbf{p}),$
here $\partial \textbf{p}$ and $\vec{n}_\textbf{p}$ sum over all faces of $\textbf{p}$, and $\mathbf{P}(T_\textbf{p})$ is the total power within $\textbf{p}$ given the temperature $T_\textbf{p}$. 
The flux terms are expressed to first order, and flux conservation across the interface between grids should be preserved. 
Take the interfaces $\sigma_1$ and $\sigma_2$ as an example; the flux is uniformly divided:
\begin{equation}
    \kappa_\textbf{p}\frac{T_\textbf{p}-T_{\sigma1}}{Lx_\textbf{p}/2} = 
    \kappa_{\textbf{q}_1}\frac{T_{\sigma1}-T_{\textbf{q}_1}}{Lx_{\textbf{q}_1}/2},\ 
    \kappa_\textbf{p}\frac{T_\textbf{p}-T_{\sigma2}}{Lx_\textbf{p}/2} = 
    \kappa_{\textbf{q}_2}\frac{T_{\sigma2}-T_{\textbf{q}_2}}{Lx_{\textbf{q}_2}/2}. \label{eq:flux}
\end{equation}
In this way, the unknown variables $T_{\sigma1},\ T_{\sigma2}$ can be expressed by linear combinations of $T_\textbf{p},\ T_{\textbf{q}_1}$, and $T_{\textbf{q}_2}$. Substitute $T_{\sigma1}$ into the flux term on $\sigma1$, it reads: $\frac{2\kappa_\textbf{p}\kappa_{\textbf{q}_1}}{\kappa_\textbf{p}+2\kappa_{\textbf{q}_1}}\frac{T_\textbf{p}-T_{\textbf{q}_1}}{Lx_\textbf{p}/2}\cdot Ly_\textbf{p}\cdot Lz_\textbf{p}$.

Similar expressions can be derived for all the neighbor grids. The discretized form of Eq.\eqref{eq:flux} for $\textbf{p}$ reads:
$\sum_{\textbf{q}\in\mathcal{N}(\textbf{p})} g_{\textbf{p},\textbf{q}}(T_\textbf{p}-T_\textbf{q})=-\mathbf{P}(T_\textbf{p})$, with $g_{\textbf{p},\textbf{q}}$ represents the coefficient term between $\textbf{p}$ and $\textbf{q}$.
Additionally, for grids near the boundary, the flux at the adiabatic interfaces is always zero. While for grids close to the ambient, for example, in the heat sink layer, the flux reads $h(T_\textbf{p})(T_\textbf{p}-T_{amb})$. 
Collect all the flux and source terms for all $\textbf{p}\in\mathcal{T}_l$; we can derive the discretized nonlinear equation at level $l$ in its general form (both conductivity and power may depend on local temperature): 
$\mathbf{G(T)T}=-\mathbf{b(T)}. \label{eq:general}$

\begin{figure}[!t]
\begin{algorithm}[H]
\caption{FAS-MG Solver}
\begin{algorithmic}[.98] \label{alg:fas}
    \REQUIRE Initial temperature $T^{(init)}$;
Convergence criteria (number of iterations $C$, maximum residual $\epsilon$, etc.)
    \ENSURE detailed solution at the finest level $\mathbf{T}_L$
    
    \STATE solve Eq.\eqref{eq:init} for initial global solution $\mathbf{T}_0^{(0)}$
    \STATE get initial solution at level $L$ by Eq.\eqref{eq:prolong}
    \WHILE {$i<C$ or $residual<\epsilon$}
        \STATE update kappa, power, and residual Eq.\eqref{eq:residual} at level L 
        \STATE calculate metrics Eq.\eqref{eq:Tr}, Eq.\eqref{eq:G} at level 1
        \STATE solve the corrected global solution by Eq.\eqref{eq:correct}
        \STATE update the finest-level solution $\mathbf{T}_L^{(i+1)}$ by Eq.\eqref{eq:update}
        \STATE smooth the finest-level solution with nonlinear Jacobi relaxations
    \ENDWHILE
\end{algorithmic}
\end{algorithm}
\end{figure}

\subsubsection{FAS-MG}
To solve the above equation, our algorithm is overviewed in Algorithm \ref{alg:fas}.
The subscript of unknown denotes the level it belongs to, and the superscript indicates the iteration count; for example, $T_l^{(i)}$ represents the solution for grids at level $l$ during the $(i)$th iteration.
Given the conductivity matrix $G(T)$ and source $b(T)$ at initial temperature $T^{(init)}$, the algorithm starts from solving:
\begin{equation}
    G_1(T^{(init)})\mathbf{T}_1^{(0)}=b_1(T^{(init)})
    \label{eq:init}
\end{equation}
for grids at level $1$ and derive the corresponding global solution $T_1^{(0)}$. It can be solved by direct solvers since the number of global grids is not large (typically tens of thousands). 
Then successive prolongation operations are conducted to interpolate the solution into the finest level $L$, deriving the initial solution $T_L^{(1)}$:
\begin{equation}
    \mathbf{T}_L^{(0)}=\mathbf{P}_{L,L-1}^{(0)}\left(
        \cdots\mathbf{P}_{1,0}^{(0)}\left(
            \mathbf{T}_0^{(0)}
            \right)\right). \label{eq:prolong}
\end{equation}
The prolongation $\mathbf{P}$ and restriction $\mathbf{R}$ operators transfer information from coarse to fine grids and vice versa.
Linear and quadratic interpolation are often utilized for prolongation \cite{BATTY201749,Teunissen2019AGM}. Still, they need to be constructed each time the local conductivity is updated when nonlinear conductivity occurs, thus being time-consuming.
In this work, we utilize the simple operation of each subgrid inheriting the value from its parent grid for prolongation, to reduce computational burden. For the restriction operator, we do not choose the common operation of simple average since it may cause many errors for grids at the interface between materials. Instead, we construct the restriction operator based on conductivity-weighted average \cite{liu2019efficient}.
To be concrete, let us take the grid $\textbf{q}$ in \figref{fig:atmesh} as an example. The prolongation operator will keep the value of $\textbf{q}_1\sim \textbf{q}_4$ the same as that of $\textbf{q}$. While the restriction operator reads $\mathbf{R}: T_\textbf{q}=\sum_{i=1-4}\kappa_{\textbf{q}_i}T_{\textbf{q}_i}/4\kappa_{\textbf{q}}$.

Then comes the FAS-MG cycles. 
Firstly, the conductivity and power are updated according to the finest-level solution $T_L^{(i)}$, and thus the conductivity matrix $G_L^{(i)}$, and residual $r_L^{(i)}$:
\begin{equation}
    r_L^{(i)} = b_L^{(i)}(\mathbf{T}_L^{(i)}) - G_L^{(i)}(\mathbf{T}_L^{(i)})\cdot \mathbf{T}_L^{(i)}
    \label{eq:residual}
\end{equation}
Then $G_0^{(i)},T_0^{(i)},$ and $r_0^{(i)}$ are transferred to level $1$ recursively:
\begin{align}
    \label{eq:Tr}
    x_0&=\mathbf{R}_{0,1}^{(i)}\left(
        \cdots\mathbf{R}_{L-1,L}^{(i)}\left(x_L\right)
        \right), \ x=\mathbf{T}^{(i)}, r^{(i)}, \\
    \mathbf{G}_0^{(i)}&=\mathbf{R}_{0,1}^{(i)}\left(
        \cdots\left(
            \mathbf{R}_{L-1,L}^{(i)}\mathbf{G}_L^{(i)}\mathbf{P}_{L,L-1}^{(i)}
            \right)\cdots
        \right)\mathbf{P}_{1,0}^{(i)}.
    \label{eq:G}
\end{align}
Traditional MG cycles calculate the solution correction with residual replacing the source term of the linear equation. However, it does not apply here due to the non-additivity of the solutions of nonlinear equations. In FAS-MG, the full equation is solved at the coarsest level instead, which reads \cite{briggs2000multigrid}:
\begin{equation}
    G_0^{(i)}\tilde{T}_0^{(i)}=
    G_0^{(i)}T_0^{(i)}+r_0^{(i)} \label{eq:correct}.
\end{equation}
Here $\tilde{T}_0^{(i)}$ is the corrected solution mitigating the high-frequency error from the finest level. Afterwards, the solution correction is prolongated until the finest level and  added to $T_L^{(i+1)}$:
\begin{equation}
    \mathbf{T}_L^{(i+1)} = \mathbf{T}_L^{(i)} + 
        \mathbf{P}_{L,L-1}^{(i)}\left(\cdots\mathbf{P}_{1,0}^{(i)}\left(\mathbf{T}_0^{(i+1)}-\mathbf{T}_0^{(i)}\right)\right)
        \label{eq:update}
\end{equation}
It is then smoothed with the smoothing operator $\mathbf{S}_1^{(1)}$ to reduce the residual. 
We employ the nonlinear Jacobi relaxation \cite{briggs2000multigrid,henson2003multigrid} to reduce the residual, as most relaxation schemes do not apply to nonlinear situations effectively. 
The iteration continues until maximum number of iterations $C$ or the convergence of solution residual, which means getting smaller than the threshold $\epsilon$. 

Although previous works \cite{iqbal2020efficient} have demonstrated the convergence of FAS-MG schemes under certain nonlinear situations, a general proof is still in lacking. 
In this work, the convergence property is assured empirically due to the physical nature of governing equation Eq. \eqref{Equ:Fourier}, which will get close to a linear equation under most thermal workloads.
Additionally, nonlinear conductivity generally does not undergo significant changes, whereas leakage power consumption can lead to thermal runaway. In such cases, our simulator address the issue by properly setting initial conditions and convergence criterion.

\subsubsection{Parallel Computing}
Our framework is well-suited for parallelization. Firstly, local refinement can be performed in parallel after global partitioning. The computation of coefficient matrices and source terms for global grids can also be parallelized. 
Additionally, we utilize parallel techniques in the iterative equation-solving stage on multiple processes.

\begin{table*}[tbh]
\caption{Performance of different simulators on linear simulation of \textbf{V100} and \textbf{3.5D-NDPs}.}
\centering
\normalsize
\resizebox{0.99\textwidth}{!}{
\begin{tabular}{c|c|cccc|cccc}
\toprule
    \multirow{2}{*}{Config.} & \multirow{2}{*}{Simulator} & \multicolumn{4}{c|}{\textbf{V100}} 
    & \multicolumn{4}{c}{\textbf{3.5D-NDPs}} \\
    & & MARE/\% & MaxE/$\SI{ }{\degreeCelsius}$ & MAE/$\SI{ }{\degreeCelsius}$ & Time/s 
    & MARE/\% & MaxE/$\SI{ }{\degreeCelsius}$ & MAE/$\SI{ }{\degreeCelsius}$ & Time/s  \\ 
    \midrule
    \multirow{4}{*}{Linear}
    &{\textbf{Icepak (gloden)}} 
    & 0 & 0 & 0 & 315 & 0 & 0 & 0 & 299 \\
    &{\textbf{HotSpot}$^\dagger$} 
    & 3.00 & 2.66 & 1.46 & 40 
    & \multicolumn{4}{c}{Cannot Handle} \\
    &{\textbf{MTA}} 
    & 1.15 & 3.08 & 0.57 & 12
    & 0.67 & 1.17 & 0.29 & {9} \\
    &{\textbf{ATSim3D}$^\dagger$} 
    & 1.34 & 1.83 & 0.69 & 8
    & \multicolumn{4}{c}{Cannot Handle} \\
    &{\textbf{Ours}} 
    & \bf{0.82} & \bf{1.36} & \bf{0.41} & \bf{5}
    & \bf{0.34} & \bf{0.95} & \bf{0.15} & \bf{3} \\ \midrule
    \multirow{3}{*}{NonLinear}
    &{\textbf{COMSOL (golden)}} 
    & 0 & 0 & 0 & 301 & 0 & 0 & 0 & 509 \\
    &{\textbf{MTA}$^\ddagger$} 
    & 1.23 & 3.98 & 0.66 & 26
    & 0.54 & {0.56} & 0.23 & {17} \\
    &{\textbf{ATSim3D}} 
    & 2.02 & 2.24 & 1.08 & 44
    &\multicolumn{4}{c}{Cannot Handle} \\
    &{\textbf{Ours}} 
    & \bf{0.83} & \bf{1.30} & \bf{0.44} & \bf{20}
    & \bf{0.49} & \bf{0.53} & \bf{0.21} & \bf{16} \\
\bottomrule

\multicolumn{10}{c}{
$^\dagger$ \small{\textbf{HotSpot} and \textbf{ATSim3D} can not handle the 3D-NDP in \textbf{3.5D-NDPs}.}
$^\ddagger$ \small{The released binary of \textbf{MTA} supports nonlinear conductivity only.}
}
\end{tabular} 
} 
\label{table:result}
\end{table*}

\begin{table*}[t]
\caption{Simulation results at the bottom of chiplet layer of systems under linear configurations and corresponding error. }
    \tiny
    \centering
    \resizebox{0.99\textwidth}{!}{
    \begin{tabular}{c|c|ccc}
    \toprule
        System & Power Map & Icepak & Ours & Error\\
        \midrule
        \textbf{V100}  
        & {\raisebox{-.4\height}{\includegraphics[height=0.5in]{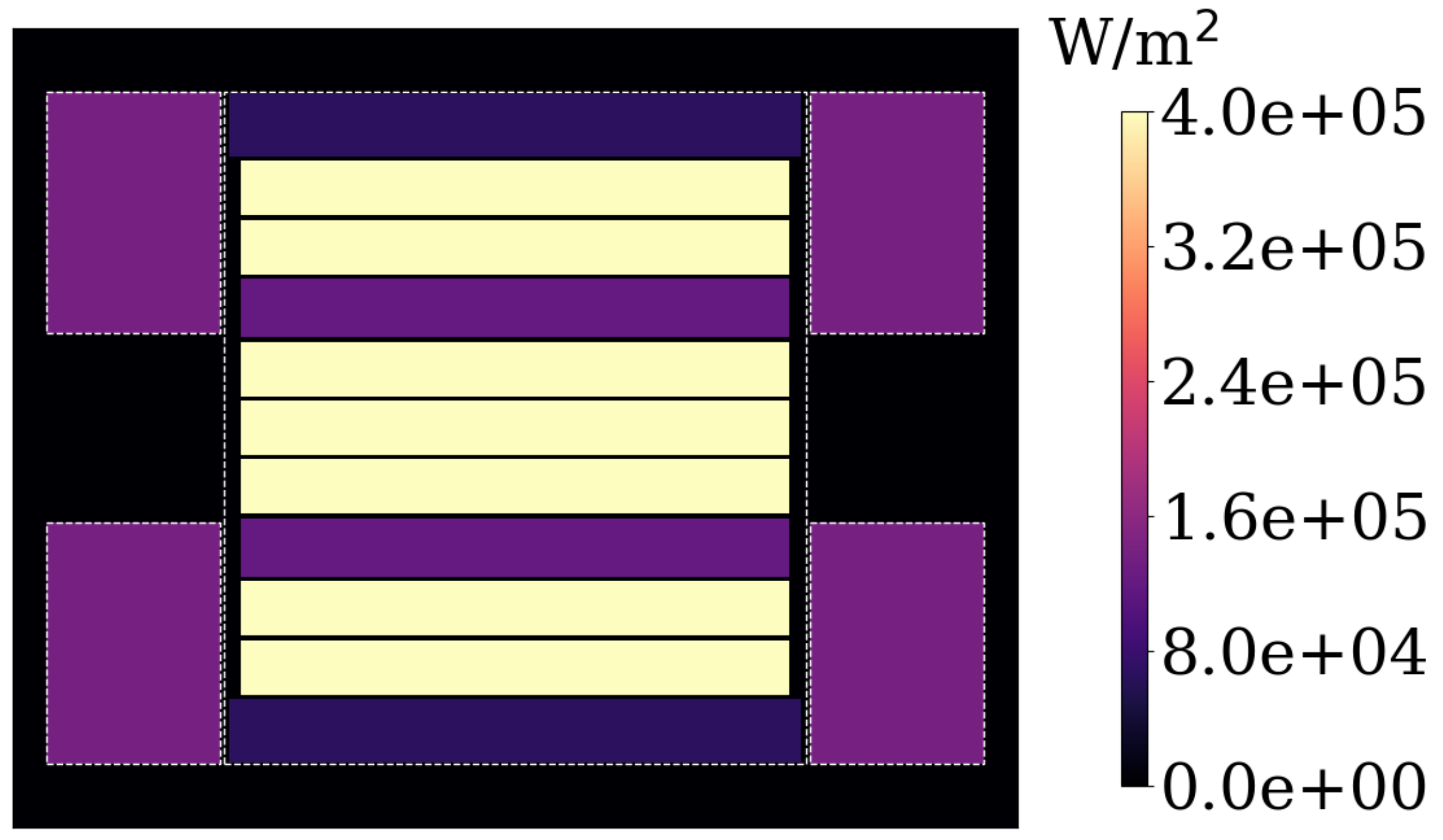}}}
        & {\raisebox{-.4\height}{\includegraphics[height=0.5in]{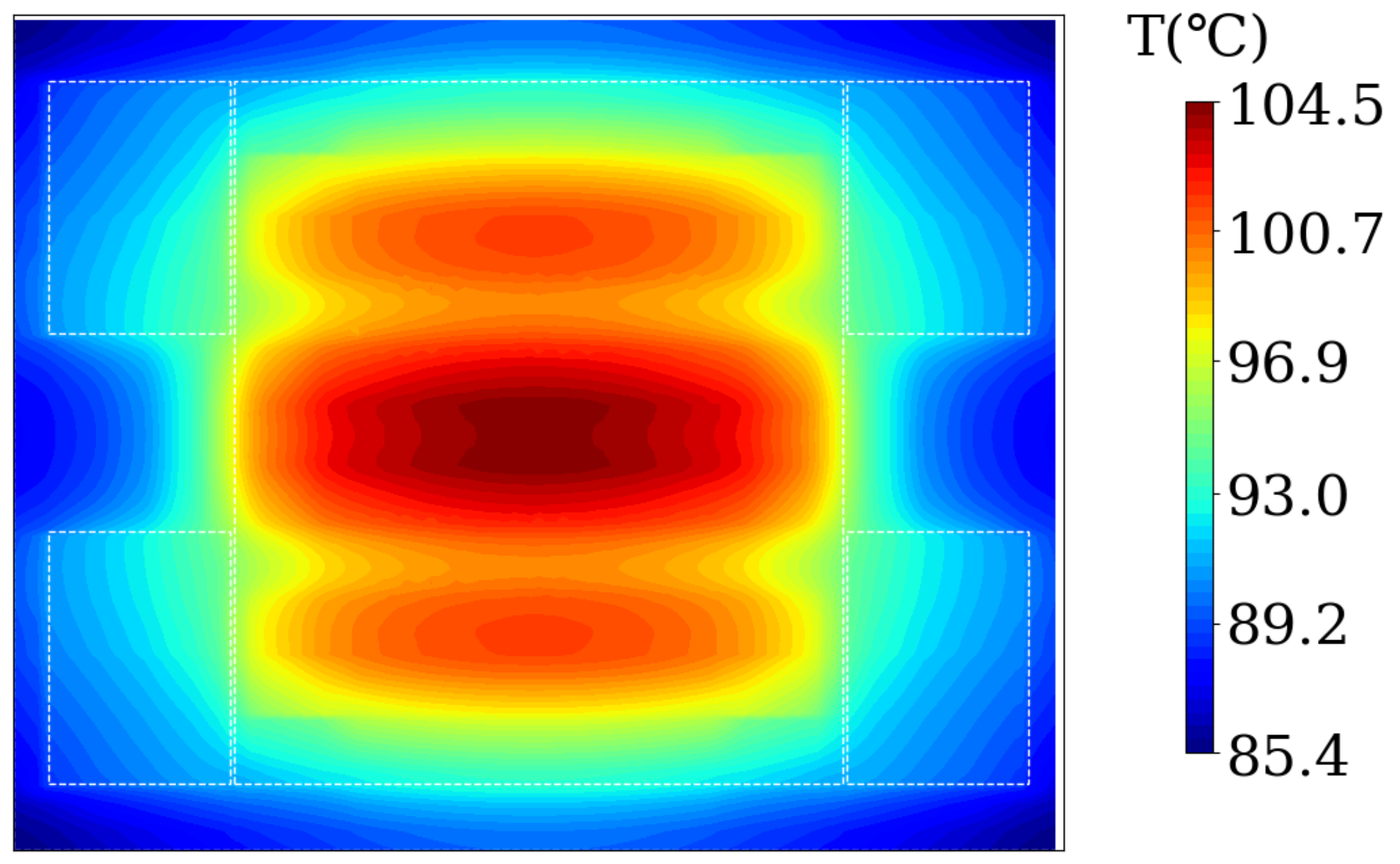}}}
        & {\raisebox{-.4\height}{\includegraphics[height=0.5in]{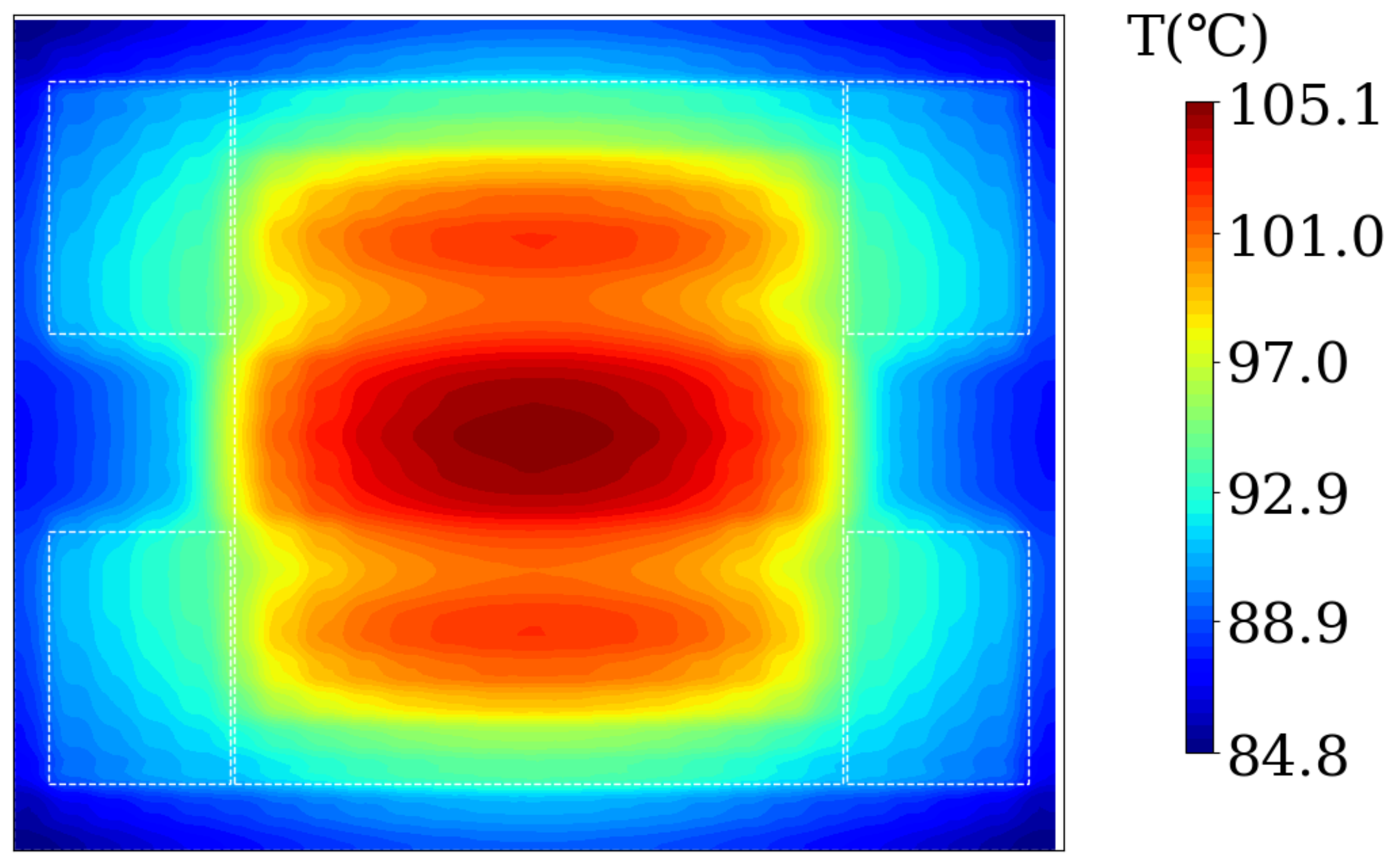}}}
        & {\raisebox{-.4\height}{\includegraphics[height=0.5in]{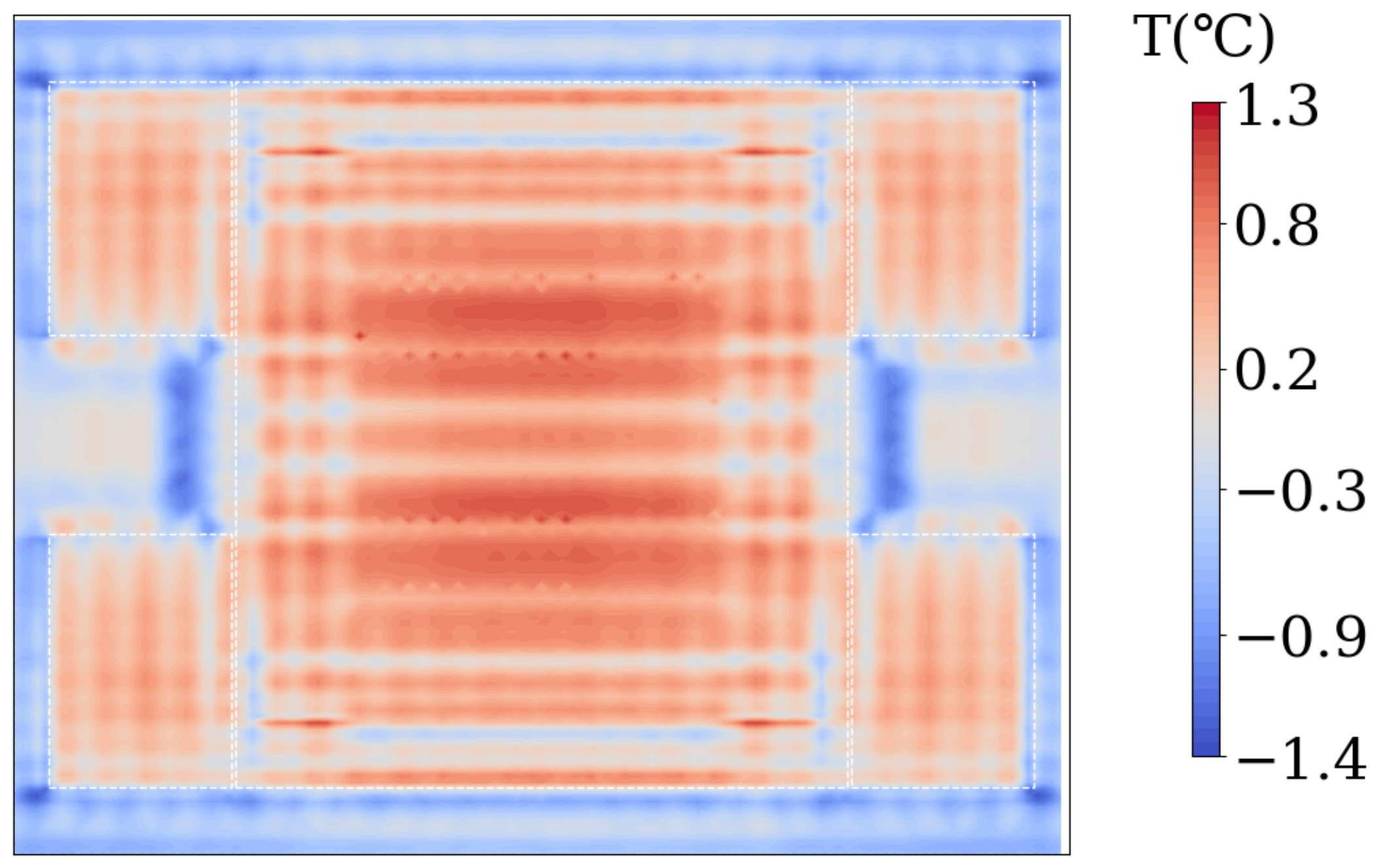}}}\\ [0.5cm] \midrule
        \textbf{3.5D-NDPs}  
        & {\raisebox{-.4\height}{\includegraphics[height=0.5in]{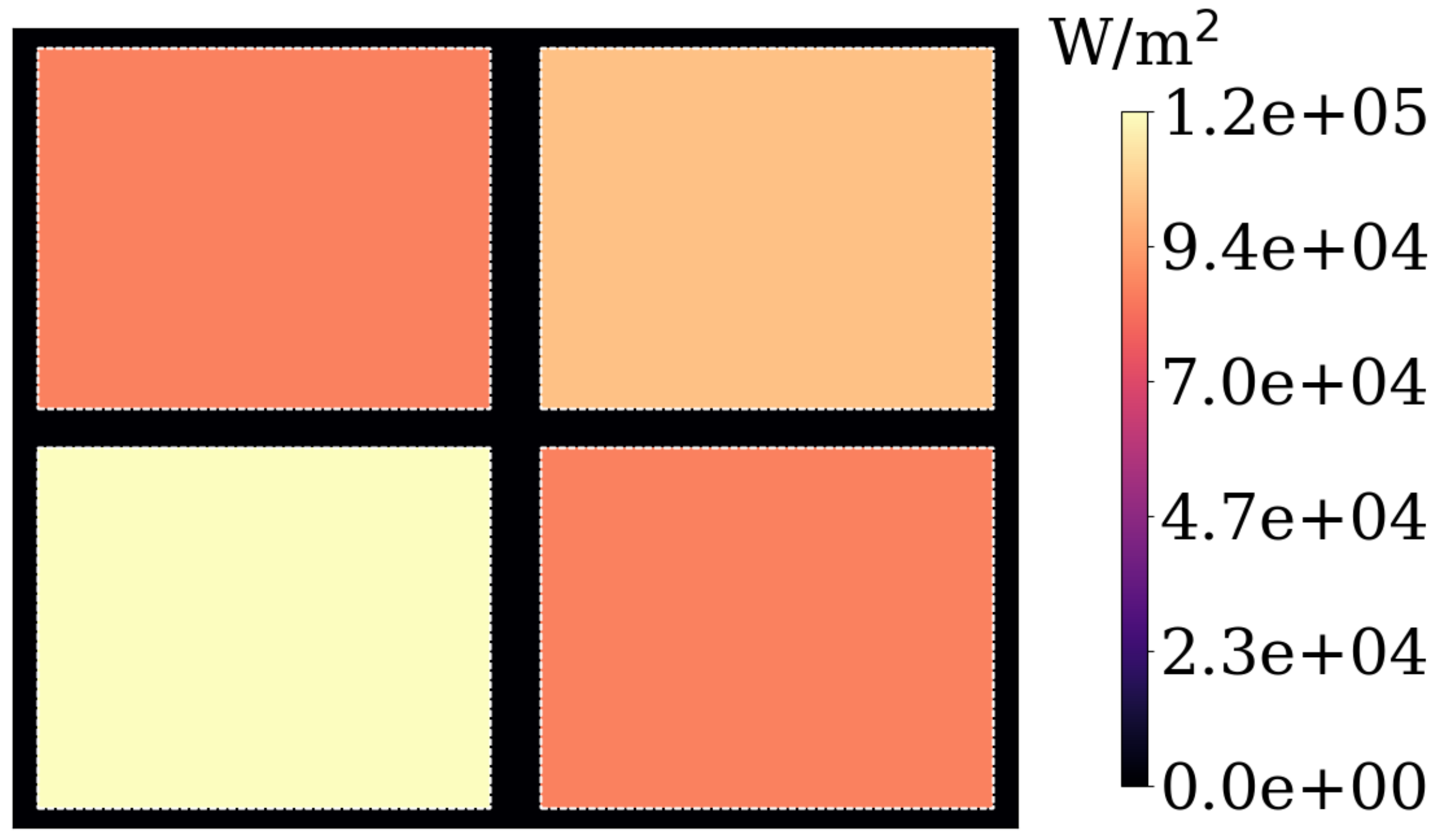}}}
        & {\raisebox{-.4\height}{\includegraphics[height=0.5in]{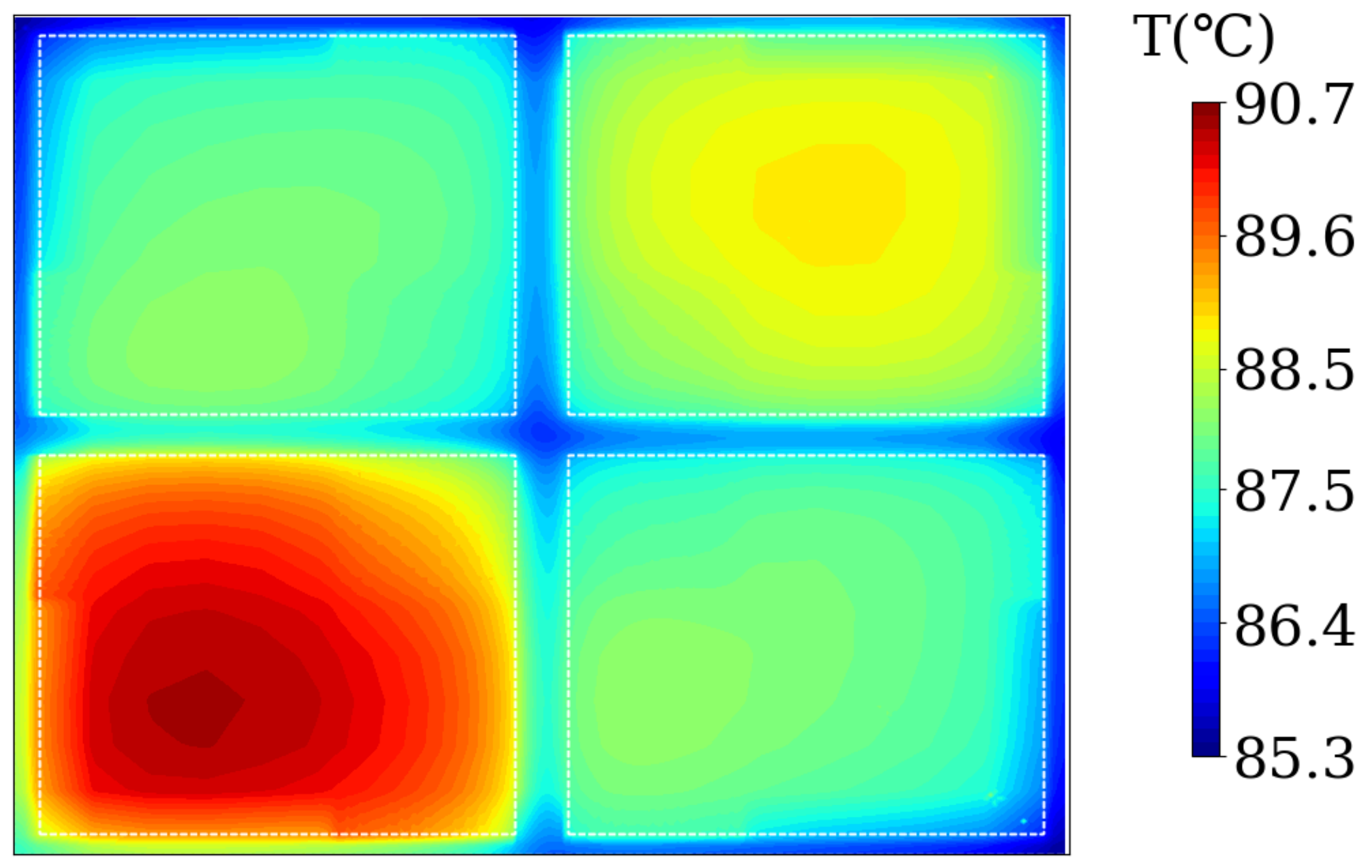}}}
        & {\raisebox{-.4\height}{\includegraphics[height=0.5in]{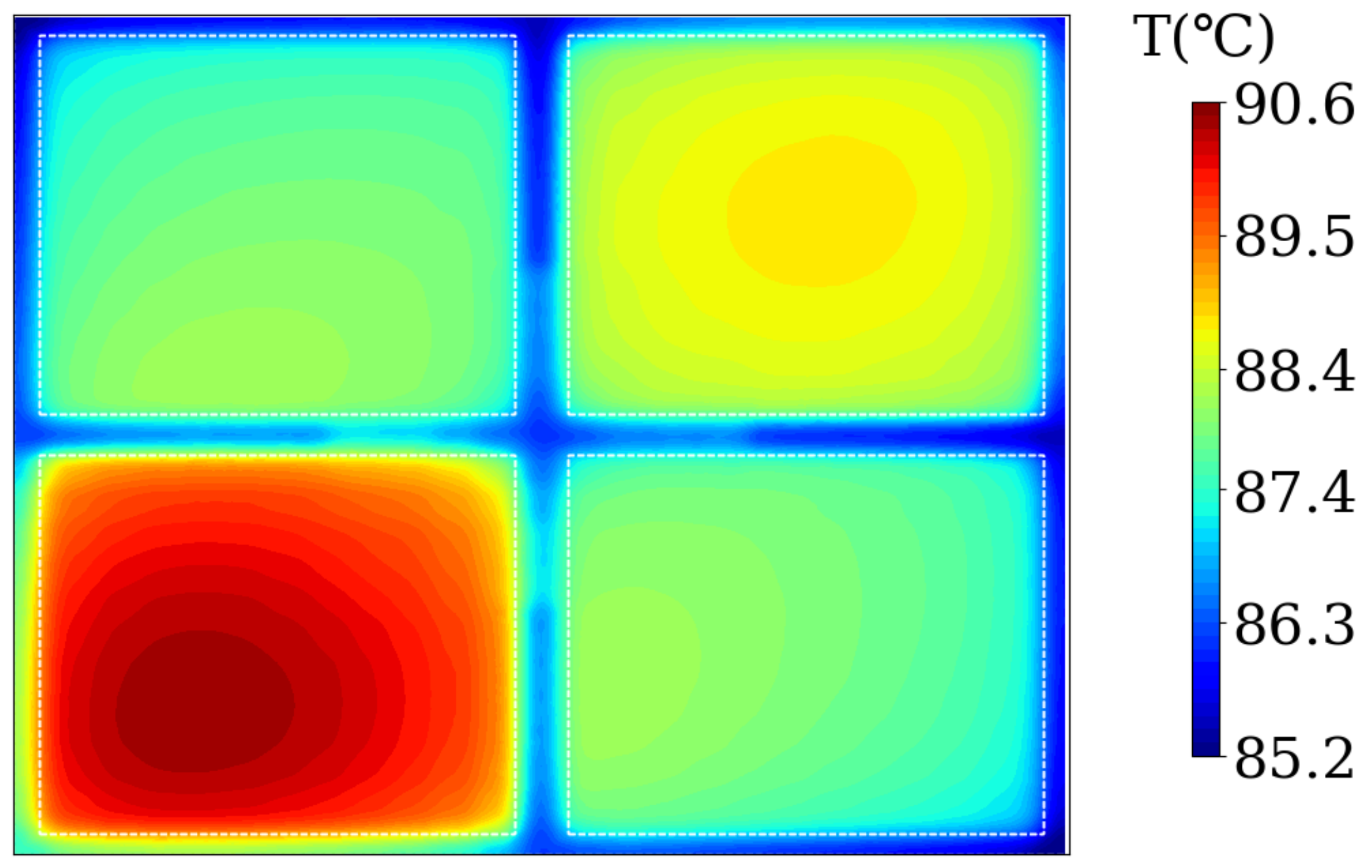}}} 
        & {\raisebox{-.4\height}{\includegraphics[height=0.5in]{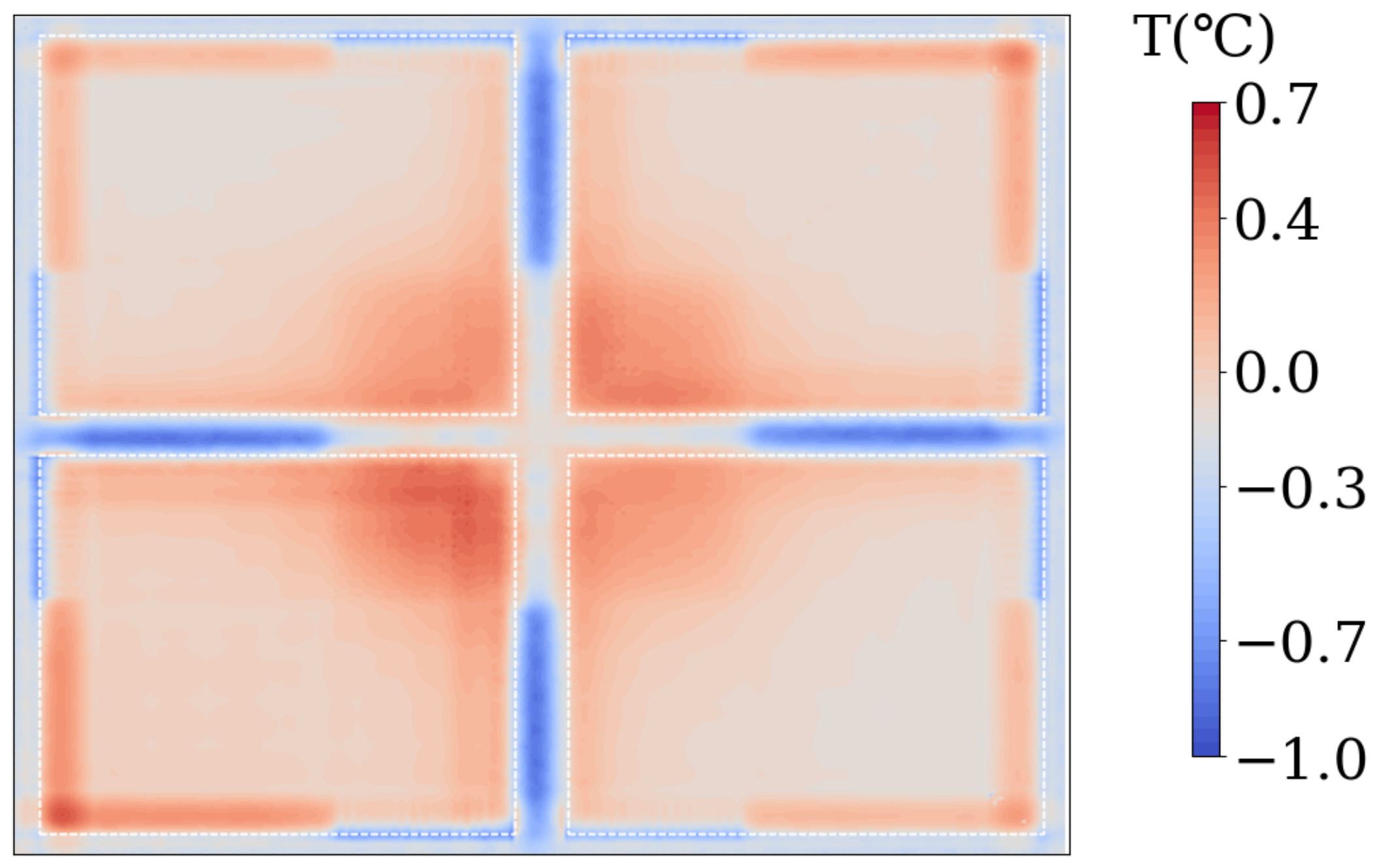}}}\\ [0.5cm]
        \bottomrule
    \end{tabular}
    }
\label{fig:linear-res}
\end{table*}

\section{Experimental Results} \label{sec:result}

\subsection{Experimental Setup}\label{sec:setup}
We arrange experiments on two chip systems, including the Nvidia V100 GPU mentioned above (labeled as \textbf{V100}) \cite{NVIDIATeslaV100}, 
and a 3.5D-IC with 3D near-data-processor (NDP) chiplets forming $2\times2$ array (labeled as \textbf{3.5D-NDPs}) modified from \cite{nie2023efficient}. The 3D-NDP chiplet encompasses 1 DRAM layer stacked on 1 logic layer.
TSV arrays composed of $31\times31$ TSVs are introduced at the center of the interposer, with the size and pitch being $100\mu m$ and $500\mu m$, respectively. 
Silicon conductivity is $140 W/mK$ under the linear configuration, and set to $\kappa_{Si}=148\cdot(300K/T)^{1.5} W/mK$ under the nonlinear one. 
Their floorplans, power maps, and material parameters are all adopted from relevant works \cite{wang2024atsim3d,wang2023efficient}. The leakage power coefficient $\beta=0.015/K@\SI{85}{\degreeCelsius}$. The HTC at the top is fixed to be $\SI{1000}{W/(m^2\cdot K)}$ and HTC at the bottom is 0 for ease of comparison with {ATSim3D}, which does not consider the secondary heat dissipation path.
Simulations are performed on a Linux server with 2 Intel Xeon 2.1GHz processors and a maximum of 40 cores. To ensure a fair comparison, appropriate mesh resolution is implemented in each experiment to guarantee convergence of the solution with mesh size and to maintain a balance between accuracy and efficiency.

\subsection{Results under Linear Configurations}

We first evaluate five simulators for the \textbf{V100} and \textbf{3.5D-NDPs} cases under the linear configuration: {Icepak} \cite{ansys}, {HotSpot} \cite{stan2003hotspot}, {MTA} \cite{ladenheim2018mta}, {ATSim3D} \cite{wang2024atsim3d}, and ours, as shown in \tabref{table:result}. 
The silicon-verified \cite{ansys_tsmc_2021} tool {Icepak} serves as the golden simulator. With an appropriate mesh setting, it features the lowest speed, followed by {HotSpot}, {MTA}, and {ATSim3D}.
The accuracy is evaluated by the mean absolute relative error (MARE), maximum error (MaxE), and mean absolute error (MAE) (defined in \secref{sec:metric}). 
{HotSpot} exhibits the largest discrepancy ($\SI{1.46}{\degreeCelsius}$ MAE for \textbf{V100}) compared to {Icepak}, followed by {ATSim3D} ($\SI{0.69}{\degreeCelsius}$ MAE for \textbf{V100}), and {MTA} ($\SI{0.57}{\degreeCelsius}$ MAE for \textbf{V100}). 
Notably, our simulator achieves the lowest error ($\SI{0.41}{\degreeCelsius}$ MAE for \textbf{V100} and $\SI{0.14}{\degreeCelsius}$ MAE for \textbf{3.5D-NDPs}), along with the highest efficiency of $80\times$ acceleration on average thanks to the multilevel grid generation strategy. 

Simulation results from {Icepak} and our simulator under the linear configuration are depicted in \tabref{fig:linear-res}, along with the corresponding power maps and error distributions. We can observe a close match between the results of ours and {Icepak}. 
Our simulator's errors arise from two sources. One is the coarse-grained error, which can be reduced by refining the global grids. The other is the lack of vertical subdivision for epoxy within the chiplet layer, which can be mitigated by treating large blocks of EMC as a dummy chiplet block. Future efforts will focus on optimizing these errors.

\begin{figure}[tbh]
    \centering
    \includegraphics[width=0.85\linewidth]{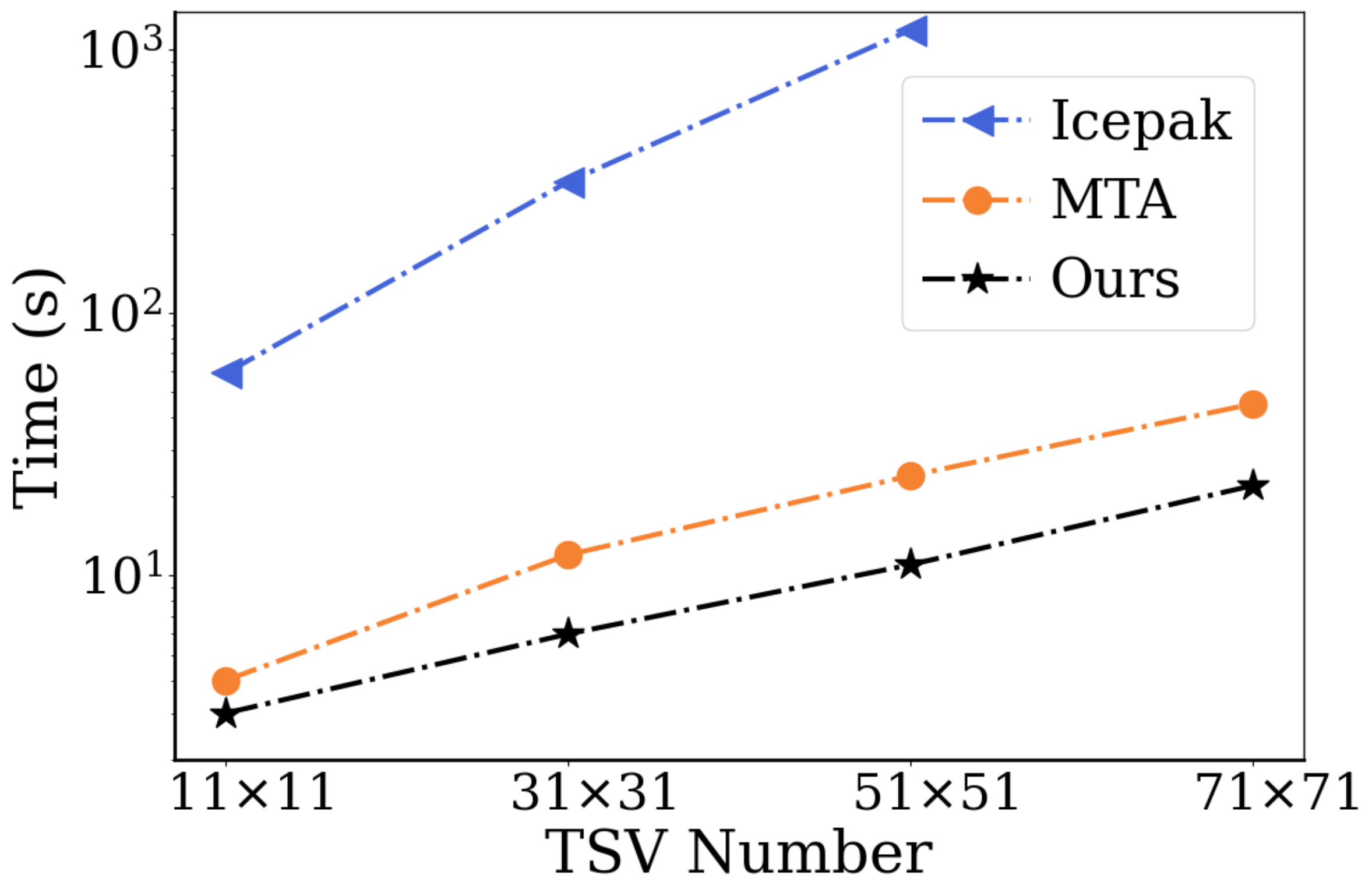}
    \caption{Runtime of simulating the \textbf{V100} with TSV arrays of different sizes under the linear configuration.}
    \label{fig:tsvs}
\end{figure}

We also test the simulators for handling large-size TSV arrays inside the \textbf{V100} system by varying the number of TSVs. 
As shown in \figref{fig:tsvs}, {Icepak} struggles to handle an array of sizes larger than 5k ($71\times71$), while our simulator exhibits better scalability and lower runtime than {MTA} concerning the growth of the number of TSVs.

\begin{figure}[tbh]
    \centering
    \includegraphics[width=0.9\linewidth]{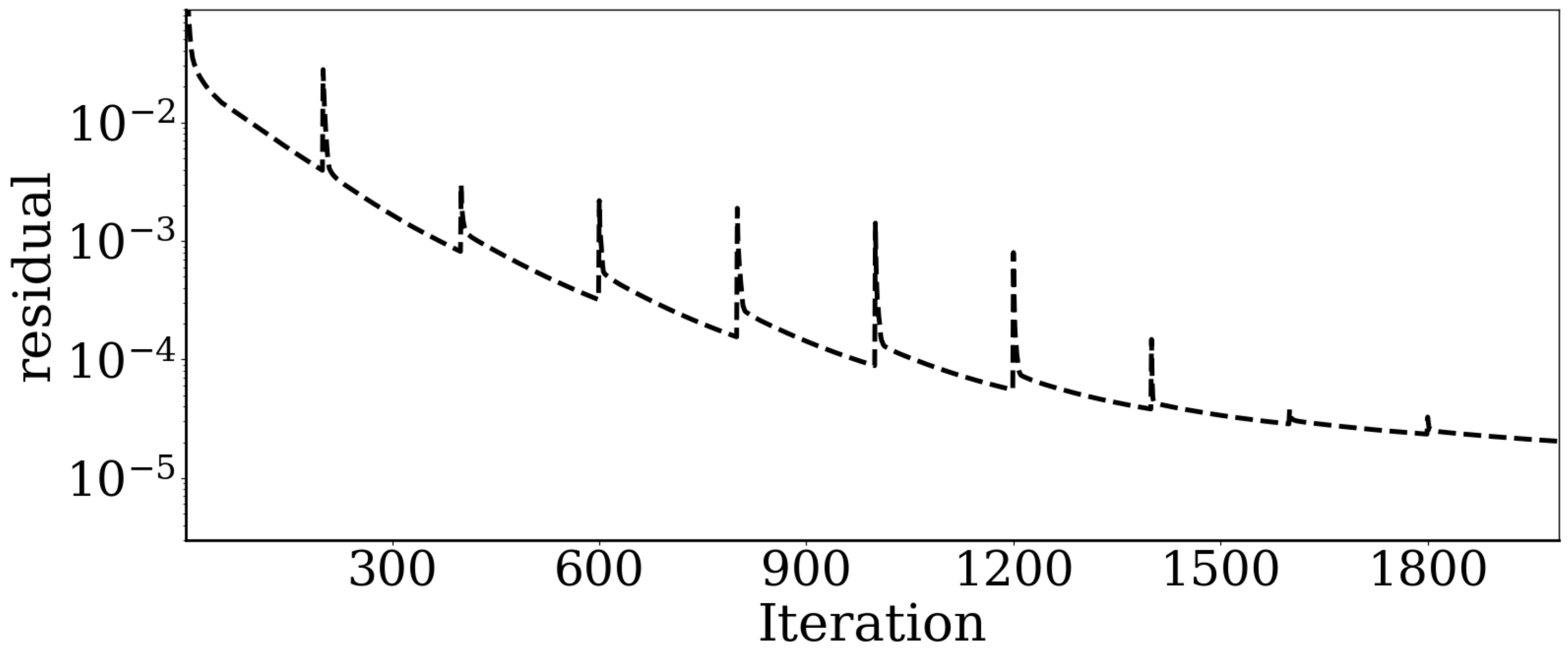}
    \caption{Convergence of residual during iterations when solving \textbf{V100} under the nonlinear configuration.}
    \label{Fig:residual}
\end{figure}

\subsection{Results under Nonlinear Configurations}

For the nonlinear simulation, as {Icepak} can only support nonlinear leakage and not nonlinear thermal conductivity, we use COMSOL as the golden simulator instead. We compare four simulators here: COMSOL (golden), {MTA}, {ATSim3D}, and ours. The released binary of {MTA} supports nonlinear conductivity only without nonlinear leakage, so we modify the leakage power according to temperature results, when running {MTA}. 

As shown in \tabref{table:result}, our simulator exhibits high accuracy ($<1\%$ MARE and $<\SI{2}{\degreeCelsius}$ MaxE) and efficiency for both \textbf{V100} and \textbf{3.5D-NDPs}, better than both {MTA} and {ATSim3D}. The runtime is much larger than that under linear configuration as the number of grids is enlarged to tackle the nonlinear conductivity.
Our simulator's errors concentrate in the high-temperature central areas, where the leakage power and conductivity are updated differently between FEM-based COMSOL and our FVM-based simulator.

\figref{Fig:residual} shows the convergence of residuals during the solving process of the \textbf{V100} system. The jumps of residual indicate the update of thermal parameters (conductivity and leakage power). 
The figure shows that the residual exhibits a rapid decrease at the beginning, followed by a gradual smoothing, indicating that our proposed method possesses excellent convergence properties, overcoming the limitation of {ATSim3D}, which is unable to continuously reduce the residual.

\newpage

\begin{table*}[tbh]
\caption{Performance of different simulators on linear simulation of \textbf{V100} and \textbf{3.5D-NDPs}.}
\centering
\normalsize
\resizebox{0.99\textwidth}{!}{
\begin{tabular}{c|c|cccc|cccc}
\toprule
    \multirow{2}{*}{Config.} & \multirow{2}{*}{Simulator} & \multicolumn{4}{c|}{\textbf{V100}} 
    & \multicolumn{4}{c}{\textbf{3.5D-NDPs}} \\
    & & MARE/\% & MaxE/$\SI{ }{\degreeCelsius}$ & MAE/$\SI{ }{\degreeCelsius}$ & Time/s 
    & MARE/\% & MaxE/$\SI{ }{\degreeCelsius}$ & MAE/$\SI{ }{\degreeCelsius}$ & Time/s  \\ 
    \midrule
    \multirow{4}{*}{Linear}
    &{\textbf{Icepak (gloden)}} 
    & 0 & 0 & 0 & 315 & 0 & 0 & 0 & 299 \\
    &{\textbf{HotSpot}$^\dagger$} 
    & 3.00 & 2.66 & 1.46 & 40 
    & \multicolumn{4}{c}{Cannot Handle} \\
    &{\textbf{MTA}} 
    & 1.15 & 3.08 & 0.57 & 12
    & 0.67 & 1.17 & 0.29 & {9} \\
    &{\textbf{ATSim3D}$^\dagger$} 
    & 1.34 & 1.83 & 0.69 & 8
    & \multicolumn{4}{c}{Cannot Handle} \\
    &{\textbf{Ours}} 
    & \bf{0.82} & \bf{1.36} & \bf{0.41} & \bf{5}
    & \bf{0.34} & \bf{0.95} & \bf{0.15} & \bf{3} \\ \midrule
\bottomrule
\end{tabular} 
} 
\end{table*}

\begin{table*}[tbh]
\caption{Performance of different simulators on linear simulation of \textbf{V100} and \textbf{3.5D-NDPs}.}
\centering
\normalsize
\resizebox{0.99\textwidth}{!}{
\begin{tabular}{c|c|cccc|cccc}
\toprule
    \multirow{2}{*}{Config.} & \multirow{2}{*}{Simulator} & \multicolumn{4}{c|}{\textbf{V100}} 
    & \multicolumn{4}{c}{\textbf{3.5D-NDPs}} \\
    & & MARE/\% & MaxE/$\SI{ }{\degreeCelsius}$ & MAE/$\SI{ }{\degreeCelsius}$ & Time/s 
    & MARE/\% & MaxE/$\SI{ }{\degreeCelsius}$ & MAE/$\SI{ }{\degreeCelsius}$ & Time/s  \\ 
    \midrule
    
    \multirow{3}{*}{NonLinear}
    &{\textbf{COMSOL (golden)}} 
    & 0 & 0 & 0 & 301 & 0 & 0 & 0 & 509 \\
    &{\textbf{MTA}$^\ddagger$} 
    & 1.23 & 3.98 & 0.66 & 26
    & 0.54 & {0.56} & 0.23 & {17} \\
    &{\textbf{ATSim3D}} 
    & 2.02 & 2.24 & 1.08 & 44
    &\multicolumn{4}{c}{Cannot Handle} \\
    &{\textbf{Ours}} 
    & \bf{0.83} & \bf{1.30} & \bf{0.44} & \bf{20}
    & \bf{0.49} & \bf{0.53} & \bf{0.21} & \bf{16} \\ \midrule
\bottomrule
\end{tabular} 
} 
\end{table*}

\section{Conclusion}
\label{sec:conclu}
As a promising approach to increase the integration of transistors, 3.5D-ICs concurrently encounter the challenges of increased power consumption and heat dissipation. 
Thermal simulation plays a vital role in the thermal design and management of 3.5D-ICs. In this work, we propose a thermal simulator designed for steady-state simulation of nonlinear and heterogeneous 3.5D-IC systems. 
We propose a new method to construct multilevel grids to capture the multiscale features and adapt the FAS multigrid method for nonlinear simulation. 
Our simulator exhibits high accuracy ($<1\%$ MARE, $<\SI{2}{\degreeCelsius}$ MaxE) and efficiency ($80\times$ acceleration) compared to commercial simulator ANSYS Icepak. 
In the future, we will generalize our methods to handle large-scale 3.5D-IC systems with more complex structures, providing a powerful means for their thermal analysis and optimization.


\newpage
\small
\bibliographystyle{IEEEtran}
\bibliography{./ref/ATSim, ./ref/Software}

@misc{comsol, 
  title={COMSOL {Multiphysics}}, 
}

@inproceedings{wang2024atsim3d,
  title={ATSim3D: Towards Accurate Thermal Simulator for Heterogeneous 3D-IC Systems Considering Nonlinear Leakage and Conductivity},
  author={Wang, Qipan and Zhu, Tianxiang and Lin, Yibo and Wang, Runsheng and Huang, Ru},
  booktitle={2024 2nd International Symposium of Electronics Design Automation (ISEDA)},
  pages={618--623},
  year={2024},
  organization={IEEE}
}

@book{briggs2000multigrid,
  title={A multigrid tutorial},
  author={Briggs, William L and Henson, Van Emden and McCormick, Steve F},
  year={2000},
  publisher={SIAM}
}

@article{smy2001transient,
  title={Transient 3D heat flow analysis for integrated circuit devices using the transmission line matrix method on a quad tree mesh},
  author={Smy, T and Walkey, D and Dew, SK},
  journal={Solid-State Electronics},
  volume={45},
  number={7},
  pages={1137--1148},
  year={2001},
  publisher={Elsevier}
}

@inproceedings{2024fastherm,
  title={FaStTherm: Fast and Stable Full-Chip Transient Thermal Predictor Considering Nonlinear Effects},
  author={Zhu, Tianxiang and Wang, Qipan and Lin, Yibo and Wang, Runsheng and Huang, Ru},
  booktitle={2024 IEEE/ACM International Conference on Computer-Aided Design (ICCAD)},
  pages={1--8},
  year={2024},
  organization={IEEE}
}

@inproceedings{2024atplace,
  title={ATPlace2.5D: Analytical Thermal-Aware Chiplet Placement Framework for Large-Scale 2.5D-IC},
  author={Wang, Qipan and Li, Xueqing and Jia, Tianyu  and Lin, Yibo and Wang, Runsheng and Huang, Ru},
  booktitle={2024 IEEE/ACM International Conference on Computer-Aided Design (ICCAD)},
  pages={1--8},
  year={2024},
  organization={IEEE}
}

@ARTICLE{10287686,
  author={Wang, Chenghan and Xu, Qinzhi and Nie, Chuanjun and Cao, He and Liu, Jianyun and Zhang, Daoqing and Li, Zhiqiang},
  journal={IEEE Transactions on Very Large Scale Integration (VLSI) Systems}, 
  title={A Multiscale Anisotropic Thermal Model of Chiplet Heterogeneous Integration System}, 
  year={2024},
  volume={32},
  number={1},
  pages={178-189},
  doi={10.1109/TVLSI.2023.3321933}}

@article{stan2003hotspot,
  title={Hotspot: A dynamic compact thermal model at the processor-architecture level},
  author={Stan, Mircea R and Skadron, Kevin and Barcella, Marco and Huang, Wei and Sankaranarayanan, Karthik and Velusamy, Sivakumar},
  journal={Microelectronics Journal},
  volume={34},
  number={12},
  pages={1153--1165},
  year={2003},
  publisher={Elsevier}
}

@article{yang2006isac,
  title={ISAC: Integrated space-and-time-adaptive chip-package thermal analysis},
  author={Yang, Yonghong and Gu, Zhenyu and Zhu, Changyun and Dick, Robert P and Shang, Li},
  journal={IEEE Transactions on Computer-Aided Design of Integrated Circuits and Systems},
  volume={26},
  number={1},
  pages={86--99},
  year={2006},
  publisher={IEEE}
}

@article{henson2003multigrid,
  title={Multigrid methods nonlinear problems: an overview},
  author={Henson, Van and others},
  journal={Computational imaging},
  volume={5016},
  pages={36--48},
  year={2003},
  publisher={SPIE}
}

@inproceedings{shukla2019overview,
  title={An overview of thermal challenges and opportunities for monolithic 3D ICs},
  author={Shukla, Prachi and Coskun, Ayse K and Pavlidis, Vasilis F and Salman, Emre},
  booktitle={Proceedings of the 2019 on Great Lakes Symposium on VLSI},
  pages={439--444},
  year={2019}
}

@inproceedings{yan2017efficient,
  title={An efficient leakage-aware thermal simulation approach for 3D-ICs using corrected linearized model and algebraic multigrid},
  author={Yan, Chao and Zhu, Hengliang and Zhou, Dian and Zeng, Xuan},
  booktitle={Design, Automation \& Test in Europe Conference \& Exhibition (DATE), 2017},
  pages={1207--1212},
  year={2017},
  organization={IEEE}
}

@article{ladenheim2018mta,
  title={The MTA: An advanced and versatile thermal simulator for integrated systems},
  author={Ladenheim, Scott and Chen, Yi-Chung and Mihajlovi{\'c}, Milan and Pavlidis, Vasilis F},
  journal={IEEE Transactions on Computer-Aided Design of Integrated Circuits and Systems},
  volume={37},
  number={12},
  pages={3123--3136},
  year={2018},
  publisher={IEEE}
}

@inproceedings{ramalingam2006accurate,
  title={Accurate thermal analysis considering nonlinear thermal conductivity},
  author={Ramalingam, Anand and Liu, Frank and Nassif, Sani R and Pan, David Z},
  booktitle={7th International Symposium on Quality Electronic Design (ISQED'06)},
  pages={6--pp},
  year={2006},
  organization={IEEE}
}

@article{nie2023efficient,
  title={Efficient transient thermal analysis of chiplet heterogeneous integration},
  author={Nie, Chuanjun and Xu, Qinzhi and Wang, Chenghan and Cao, He and Liu, Jianyun and Li, Zhiqiang},
  journal={Applied Thermal Engineering},
  volume={229},
  pages={120609},
  year={2023},
  publisher={Elsevier}
}

@article{terraneo20213d,
  title={3D-ICE 3.0: efficient nonlinear MPSoC thermal simulation with pluggable heat sink models},
  author={Terraneo, Federico and Leva, Alberto and Fornaciari, William and Zapater, Marina and Atienza, David},
  journal={IEEE Transactions on Computer-Aided Design of Integrated Circuits and Systems},
  volume={41},
  number={4},
  pages={1062--1075},
  year={2021},
  publisher={IEEE}
}

@article{liu2019efficient,
  title={An efficient strategy for large scale 3D simulation of heterogeneous materials to predict effective thermal conductivity},
  author={Liu, Xiaodong and R{\'e}thor{\'e}, Julien and Baietto, Marie-Christine and Sainsot, Philippe and Lubrecht, Antonius Adrianus},
  journal={Computational materials science},
  volume={166},
  pages={265--275},
  year={2019},
  publisher={Elsevier}
}

@article{iqbal2020efficient,
  title={An efficient nonlinear multigrid scheme for 2D boundary value problems},
  author={Iqbal, Sehar and Zegeling, Paul Andries},
  journal={Applied Mathematics and Computation},
  volume={372},
  pages={124898},
  year={2020},
  publisher={Elsevier}
}

@misc{NVIDIATeslaV100,
  author = {NVIDIA Corporation},
  title = {{NVIDIA Tesla V100 GPU Architecture White Paper}},
  howpublished = {\url{http://images.nvidia.com/content/volta-architecture/pdf/volta-architecture-whitepaper.pdf}},
  year = {2017},
}

@article{wang2023efficient,
  title={An efficient thermal model of chiplet heterogeneous integration system for steady-state temperature prediction},
  author={Wang, Chenghan and Xu, Qinzhi and Nie, Chuanjun and Cao, He and Liu, Jianyun and Li, Zhiqiang},
  journal={Microelectronics Reliability},
  volume={146},
  pages={115006},
  year={2023},
  publisher={Elsevier}
}

@article{Teunissen2019AGM,
  title={A geometric multigrid library for quadtree/octree AMR grids coupled to MPI-AMRVAC},
  author={Jannis Teunissen and Rony Keppens},
  journal={Comput. Phys. Commun.},
  year={2019},
  volume={245},
}

@misc{ansys_tsmc_2021,
  author = {{ANSYS}},
  title = {{ANSYS collaborates with TSMC to deliver thermal analysis solution for 3D IC designs}},
  year = {2021-10-27}
}

@article{BATTY201749,
	author = {Christopher Batty},
	doi = {https://doi.org/10.1016/j.jcp.2016.11.035},
	issn = {0021-9991},
	journal = {Journal of Computational Physics},
	pages = {49-72},
	title = {A cell-centred finite volume method for the Poisson problem on non-graded quadtrees with second order accurate gradients},
	volume = {331},
	year = {2017},
  }

@INPROCEEDINGS{10195617,
  author={Lee, Ilbok and Nam, Soohyun and Kim, Sungeun and Shin, Sangho and Kim, Younglyong and Seo, Sun-Kyoung and Yu, Hae Jung and Kim, Dae-Woo},
  booktitle={2023 IEEE 73rd Electronic Components and Technology Conference (ECTC)}, 
  title={Extremely Large 3.5D Heterogeneous Integration for the Next-Generation Packaging Technology}, 
  year={2023},
  volume={},
  number={},
  pages={893-898},
  doi={10.1109/ECTC51909.2023.00154}
}

@article{lau2025current,
  title={Current Advances and Outlooks in Hybrid Bonding},
  author={Lau, John H},
  journal={IEEE Transactions on Components, Packaging and Manufacturing Technology},
  year={2025},
  publisher={IEEE}
}

@misc{ansys,
  title        = {ANSYS {ICEPAK}},
  howpublished = "\url{http://www.ansys.com}"
}

@string{date     = "Proc.~DATE"}

@string{ectc     = "Proc.~ECTC"}

@string{ics      = "Proc.~ICS"}

@string{iccad    = "Proc.~ICCAD"}

@string{isqed    = "Proc.~ISQED"}

@string{spie     = "Proc.~SPIE"}

@string{tsmc      = "IEEE Transactions on Systems, Man, and Cybernetics: Systems"}

@string{test     = "Test"}

@string{science   = "Science"}
\end{document}